\pdfoutput=1
\documentclass[twocolumn,10pt]{asme2ej}

\usepackage{epsfig}
\usepackage{gensymb}
\usepackage{nomencl}
\usepackage{natbib}
\usepackage{graphicx}
\usepackage{txfonts}
\usepackage{array}
\usepackage{fancyhdr}
\usepackage{float}
\usepackage{caption}
\usepackage{subcaption}

\fancypagestyle{newstyle}{
\fancyhf{} % clear all header and footer fields
\cfoot{\thepage}
}
\fancypagestyle{newstyle2}{
\fancyhf{} % clear all header and footer fields
\lfoot{\it Preprint submitted to ASME. }
\rfoot{ \it May 1, 2014}
}

\title{Numerical Investigation of Effects of Compound Angle and Length to Diameter Ratio on Adiabatic Film Cooling Effectiveness }

\author{Vidit Sharma, Ashish Garg\thanks{Corresponding Author\lq s email: ashish@gateaerospaceforum.com } 
    \affiliation{
	 {\it GATE Aerospace Forum Educational Services, New Delhi - 110059, India}
    }	
}

\begin{document}
%\clearpage

\maketitle  
\thispagestyle{newstyle2}

%%%%%%%%%%%%%%%%%%%%%%%%%%%%%%%%%%%%%%%%%%%%%%%%%%%%%%%%%%%%%%%%%%%%%%
\begin{abstract}
{\it A modification has been done in the normal injection hole of $35\degree$, by injecting the cold  fluid at different angles(compound angle) in lateral direction, providing a significant change in the shape of holes which later we found in our numerical investigation giving good quality of effectiveness in cooling. Different L/D ratios are also studied for each compound angle. The numerical simulation is performed based on Reynolds Averaged Navier-Stokes(RANS) equations with k-$\epsilon$ turbulence model by using Fluent(Commercial Software). Adiabatic Film Cooling Effectiveness has been studied for compound angles of ($0\degree$, $30\degree$, $45\degree$ and $60\degree$) and L/D ratios of (1, 2, 3 and 4)  on a hole of 6mm diameter with blowing ratio 0.5. The findings are obtained from the results, concludes that the trend of laterally averaged adiabatic effectiveness is the function of  L/D ratio and compound angle.  \par
}
\end{abstract}
%\clearpage
%\begin{keywords}
%{ \it Keywords : film cooling, turbine cooling, computational simulation on turbine blades } 
%\end{keywords}
%

%%%%%%%%%%%%%%%%%%%%%%%%%%%%%%%%%%%%%%%%%%%%%%%%%%%%%%%%%%%%%%%%%%%%%%
\section*{Nomenclature}
  
\[\begin{array}{lp{0.8\linewidth}}
        D  & diameter of the film cooling hole    \\
        DR &density ratio of jet to free stream = $\frac{\rho_c}{\rho_h}$                 \\
        L & hole length \\
        M & blowing ratio of jet to free stream = $\frac{\rho_c U_c}{\rho_h U_H}$                 \\
         T_H             & hot free stream absolute temperature, K \\
         T_P             & temperature at any plane, K \\
         Pr	&Prandtl Number           \\
         U_H & velocity, m/s \\
         x	&streamwise distance, m\\
y	&spanwise distance, m \\
y+	&non-dimensional wall distance for a wall-bounded flow
      \end{array}
\]

\section*{Greek}
%\vskip .1em 
%\lineskip 1em 
\[\begin{array}{lp{0.8\linewidth}}
 \alpha &injection angle, in degree        \\
         \rho         &  density     \\
         \mu	&dynamic viscosity, Pa-s \\
\beta	&Compound angle, in degree \\
\eta	&adiabatic film cooling effectiveness, $\frac{T_H-T_W}{T_H-T_C}$ \\
\eta_p	& normalised temperature,  $\frac{T_H-T_P}{T_H-T_C}$ \\
\bar{\eta} &laterally averaged adiabatic film cooling effectiveness
      \end{array}
\]

\section*{Subscript}
%\vskip .1em 
\[\begin{array}{lp{0.8\linewidth}}

H	&Mainstream \\
avg	&lateral average \\
C	&coolant\\
W	&Wall \\
P   &any Plane of consideration
 \end{array}
\]
%\makenomenclature
%\nomenclature{D}{diameter of the film cooling hole} %
%\nomenclature{DR}{density ratio of jet to free stream = $\frac{\rho_c}{\rho_h}$ } %
%\nomenclature{$\alpha$}{injection angle, in degree} %
%\nomenclature{$\psi$}{denotes angular coordinates,direction component velocity} %
%\nomenclature{$\nu$}{radial or y direction component velocity}
%\nomenclature {M}{Mach number}
%\nomenclature {$\gamma$}{ratio of specific heat}
%\nomenclature{$\rho$}{density}
%\printnomenclature
%\end{nomenclature}
%%%%%%%%%%%%%%%%%%%%%%%%%%%%%%%%%%%%%%%%%%%%%%%%%%%%%%%%%%%%%%%%%%%%%%

\section*{Introduction}
Gas turbines has been used in aviation and many other industries for power generation. To achieve efficient power generation from these machines higher inlet temperature for turbine is needed. However, the operating temperature cannot be increased because of the material's limiting temperature. Very high temperature degrade the blades and other parts of the turbine. Several cooling techniques are in use to protect the gas turbine parts exposed to extreme temperature. There are several techniques to cool gas turbine blades out of those, inner lining Film cooling is one of them. For several decades film cooling has been used widely for cooling of gas turbines. Film cooling hole shapes have a large effect on the performance on film cooling effectiveness. However, shapes of the holes are limited due to manufacturing, durability, maintenance, etc. These days due to advancement in manufacturing techniques people are getting more towards the shaped holes to increase the effectiveness in cooling.\par 
Several studies has been done on film cooling and several factors came in front which effects the film cooling effectiveness. These factor can be studied further to improve the film cooling such as hole diameter, hole shape, injection angle, compound angle, orientation angle, Reynolds number, blowing ratio, density ratio, position and turbulent intensity of hot stream.\par
Several computational and experimental studies has been done on film cooling is reviewed in this section. Mehendale et al. [1] studied the film cooling effectiveness experimentally over the leading edge with variation of Reynolds number. The experiment is performed for the Reynolds numbers 25000, 40000 and 100000 with other parameters. It was seen that with increase in Reynolds number of mainstream gives better film cooling effectiveness and this similar trend was also concluded by Ou et al. [2] for leading edge film cooling.\par
Kohli and Bogard [3] performed an experimental study on film cooling with two injection angles and they found that $35\degree$ injection angle provide higher center-line and laterally averaged adiabatic effectiveness than $55\degree$ injection angle. Lampard and Foster [4] studied experimentally and concluded that lower injection angles gives higher film cooling effectiveness at lower blowing ratios and higher injection angles gives higher film cooling effectiveness at higher blowing ratios.\par
Blowing ratio effect is provided by Cho et al [5] through experimental study says that at lower blowing ratio M=0.5 film cooling effectiveness is almost same for all holes and as blowing ratio rate increases film cooling effectiveness decreases. Talib et al [6] also concluded through experimental study that at $45\degree$ injection angle film cooling effectiveness is better for blowing ratio 0.64 and 0.50 then blowing ratio 0.94 and 0.83.  Wright et al. [7] also carried out an experimental study with $35\degree$ injection angle and in this case blowing ratio various from 0.25 to 2.0 and density ratio various from 1 to 1.4 and from this experimental study it has been found that film cooling effectiveness increases with increase in density ratio and decreases with increase in blowing ratio. Bogard and Crawford [8] studied effect of density ratio on film cooling effectiveness and concluded that increase in blowing ratio with decrease in density ratio lead to reduce lateral spreading of the coolant on the plate and thus reduce laterally averaged effectiveness. \par
Ethridge et al. [9] carried out an experimental study on film cooling effectiveness on suction side of the first stage turbine vane with discrete injection holes. In this case density ratio various from 1.1 to 1.6 and blowing ratio various from 0.2 to 1.5. It has been found that with blowing ratio less than 0.5 there is a sharp decrease in the film cooling effectiveness with small density ratio and this trend was not observed in large density ratios. \par
Nasir and Ekkad [10] studied experimentally the effect of compound hole with injection angle on film cooling effectiveness and concluded that $60\degree$ lateral angle compound hole with higher injection angle $55\degree$ will give lower film cooling effectiveness as compared to $35\degree$ injection angle due to the lateral angle effect of the hole. Jung and Lee [11] have study the effects of orientation angles (compound holes) and velocity ratios on the film cooling effectiveness experimentally. In this study orientation angles were $0\degree$, $30\degree$, $60\degree$ and $90\degree$ and velocity ratios were 0.5, 1.0, and 2.0. From this study it has been concluded that orientation angles effect on film cooling effectiveness is very dependent on velocity ratios at small velocity ratio 0.5 the increase in averaged film cooling effectiveness with increase in orientation angle is very less, the increase in averaged film cooling effectiveness at higher velocity ratio 2.0 with increase in orientation angle is large. In a thesis by Joshua P. F. Grizzle [12] an experimental study has been studied for upstream steps for injection and compound angle. In this study injection angle was $35\degree$ and compound angle was $45\degree$ and it has been concluded that averaged film cooling effectiveness is more for $35\degree$ injection angle and $45\degree$ compound angles as compared to $35\degree$ injection angle at same blowing ratio. \par
Baheri et al. [13] studied the effect of length to diameter ratio with cylindrical and trenched hole with two blowing ratio 0.6 and 1.25 and with injection angle $35\degree$. From this study it has been concluded that effect of injection angles span wise and averaged film cooling effectiveness are increased with respect to length to diameter ratio. \par
From the literature we found  that film cooling effectiveness is strongly depend upon injection angle, compound angle, orientation angle, density ratio, blowing ratio, length to diameter ratio. However effect of length to diameter ratio on compound angle is not explored. In this paper, we have studied the effect of compound angle with different length to diameter ratio at blowing ratio 0.5 with Reynolds number 38244. 
The analysis has been done using commercial CFD package which explore and resolve the three dimensional nature of flow over the flat plate to be cooled.
%%%%%%%%%%%%%%%%%%%%%%%%%%%%%%%%%%%%%%%%%%%%%%%%%%%%%%%%%%%%%%%%%%%%%%
\section*{Problem Description}

Comparison of adiabatic effectiveness for different compound angles and different L/D ratios is studied in this paper. To validate our CFD model, and results of simulation, we have considered the geometry used by Sangkwon and Tom S. [14] as shown in Fig. 1. and experimental result on it by  Kohli and Bogard [3]. \par
In Fig. 1. , Hole diameter of 12.7mm is used with length of 2.8D and with same operating conditions maintained by Kohli and Bogard [3] and Sangkwon and Tom S. [14].\par 
\begin{figure}[b]
\scalebox{1}{\includegraphics[width=3.25in]{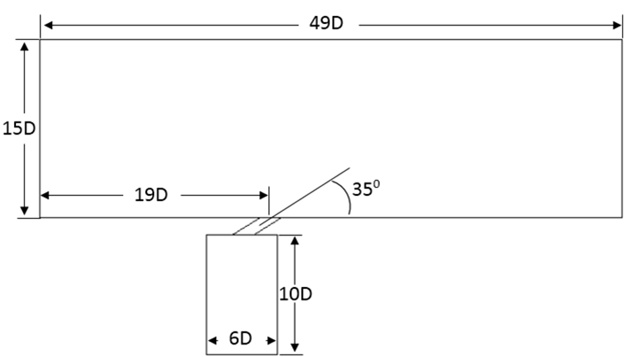}}
  %\centerline{\includegraphics{Fig6(a).eps}}% Images in 100% size
  \caption{ Diagram of the film cooling configuration studied in this paper}
\label{fig:f1}
\end{figure}

Further parametric study was done with hole diameter of 6mm with the configuration shown in Fig. 1-2. Row of holes shown in Fig. 2(a) shows that the distance between the hole centers is 3D in spanwise direction. Injection angle of the hole is $\alpha$=$35\degree$ with respect to the mainstream flow as shown in the Fig. 2(b). Compound angle $\beta$ is with respect to the y-axis which can be seen in Fig. 2(c). Different length of the hole is also study in form of a non-dimensional quantity L/D ratio. The L/D ratios studied are 1,2,3 and 4. The hole is connected to the plenum which drives the cold air through the hole over the plate. The investigation has been done on compound angle $\beta$ = $0\degree$, $30\degree$, $45\degree$ and $60\degree$.  \par

\begin{figure}
\scalebox{1}{\includegraphics[width=3.25in]{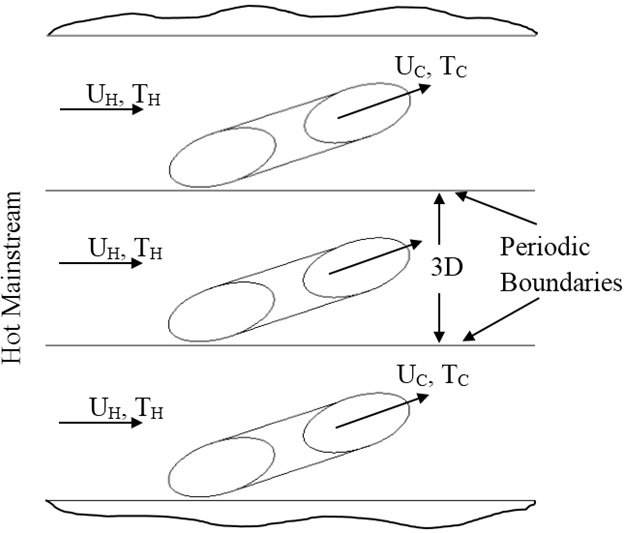}}
  %\centerline{\includegraphics{Fig6(a).eps}}% Images in 100% size
  %\caption{(a) Compound hole configuration on plate }
  \begin{center}
  (a)\\
  \end{center}
\label{fig:f2a}
\end{figure} 

\begin{figure}
\scalebox{1}{\includegraphics[width=3.25in]{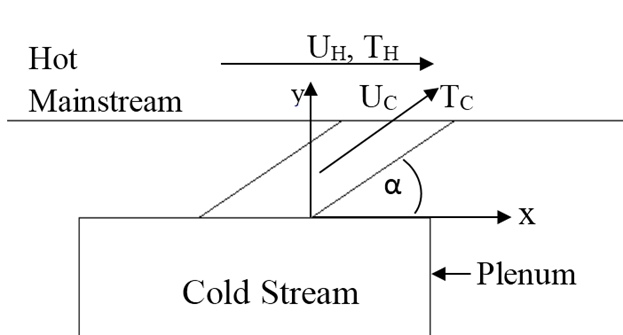}}
  %\centerline{\includegraphics{Fig6(a).eps}}% Images in 100% size
%\caption{Injection angle ?}
\begin{center}
  (b)\\
  \end{center}
\label{fig:f2b}
\end{figure} 

\begin{figure}
\scalebox{1}{\includegraphics[width=3.25in]{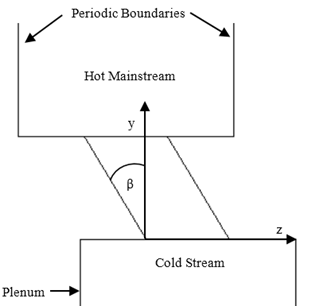}}
  %\centerline{\includegraphics{Fig6(a).eps}}% Images in 100% size
  \begin{center}
  (c)\\
  \end{center}
  \caption{ (a) Compound hole configuration on plate; (b) Injection angle $\alpha$; (c) Compound angle $\beta$ }
\label{fig:f2c}
\end{figure}

Fig. 3 shows the geometry for normal injection hole with $\alpha$ = $35\degree$ and L/D=2 and Fig. 4 shows the geometry for compound hole with $\alpha$ = $35\degree$, $\beta$ = $45\degree$ and L/D=2. From the Fig. 5(a)-(b) it can be seen that how the compound angle changes the shape of the hole with respect to the normal hole shape. \par

\begin{figure}
\scalebox{1}{\includegraphics[width=3.25in]{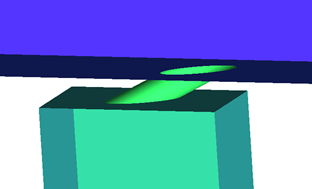}}
  %\centerline{\includegraphics{Fig6(a).eps}}% Images in 100% size
  \caption{ Geometry of compound hole $\alpha$ = $35\degree$, $\beta$ = $45\degree$ and L/D=2 }
\label{fig:f3}
\end{figure} 

\begin{figure}
\scalebox{1}{\includegraphics[width=3.25in]{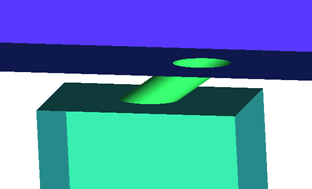}}
  %\centerline{\includegraphics{Fig6(a).eps}}% Images in 100% size
  \caption{ Geometry of normal injection hole $\alpha$ = $35\degree$ and L/D=2 }
\label{fig:f4}
\end{figure} 

\begin{figure}
\scalebox{1}{\includegraphics[width=3.25in]{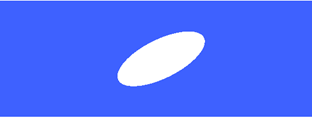}}
  %\centerline{\includegraphics{Fig6(a).eps}}% Images in 100% size
 % \caption{ (c) Compound angle ?}
 \begin{center}
  (a)\\
  \end{center}
\label{fig:f5a}
\end{figure} 

\begin{figure}
\scalebox{1}{\includegraphics[width=3.25in]{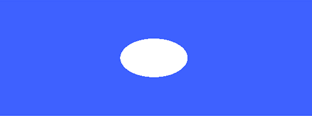}}
  %\centerline{\includegraphics{Fig6(a).eps}}% Images in 100% size
  \begin{center}
  (b)\\
  \end{center}
  \caption{ (a) Shape of compound hole on plate with $\alpha$ = $35\degree$, $\beta$ = $45\degree$ and L/D=2 ; (b) Shape of normal injection hole on plate with $\alpha$ = $35\degree$ and L/D=2  }
\label{fig:f5b}
\end{figure} 

Mainstream flow of air over the flat plate has a temperature of 300k. And the secondary cold flow coming out of the hole through the plenum is at temperature 188k. Mainstream flow over the plate is at a velocity of 20m/s. While at the exit of the hole, velocity of cold air is 6.25m/s maintaining a blowing ratio of 0.5. As the cold air interacts with the mainstream air, it spreads on the plate and reducing the temperature of plate. Thus, the adiabatic film cooling effectiveness can be calculated as follows \par
\begin{equation}
\eta = \frac{T_H-T_W}{T_H-T_C} 
\label{eq_ASME}
\end{equation}
For compound hole it is not possible to compare centerline effectiveness as the cold stream spread on the flat plate is not symmetric in lateral direction. Therefore, only laterally averaged adiabatic film cooling effectiveness is obtained for comparison which is done by extracting wall temperature from each node of surface grid in lateral direction and then averaged. \par
To obtain the numerical solution, we have specified the given boundary conditions on geometry and used  Fluent 14.0.0 for our simulation. These boundary conditions for the computational domain are as in Fig. 6. At the inlets of the mainstream all velocity components, and temperatures are specified. The walls of the main duct as well as secondary duct are non-slip walls with adiabatic condition. Left and right side of mainstream duct and plenum is periodic boundary conditions. Near wall turbulence is handled using enhanced wall model of Realizable k-$\epsilon$. The outlet boundary is treated as pressure outlet with zero gauge pressure.

\begin{figure}
\scalebox{1}{\includegraphics[width=3.25in]{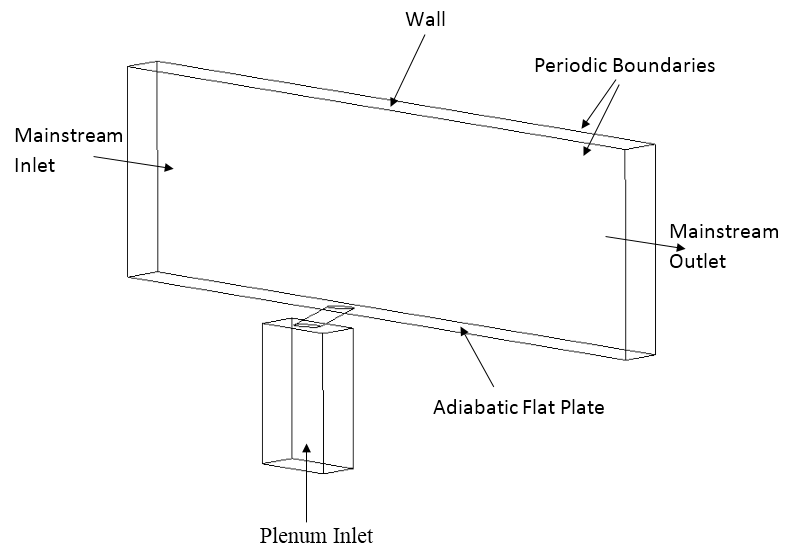}}
  %\centerline{\includegraphics{Fig6(a).eps}}% Images in 100% size
  \caption{ Boundary conditions for computation domain}
\label{fig:f6}
\end{figure}

%\begin{figure}(100,100)(40,250)
%\includegraphics[scale=1]{Fig1.png}
%\end{figure}
%
%\begin{figure} 
%\centerline{\fig{Fig6(a).eps,width=3.34in}}
%\caption{Hi Baby.  ASME sets figures captions in 8pt, Helvetica Bold.}
%\label{fig_example1.png}
%\end{figure}
%\begin{picture}
%\includegraphics[scale=0.15]{Fig1.png}
%\end{picture}\\

%%%%%%%%%%%%%%%%%%%%%%%%%%%%%%%%%%%%%%%%%%%%%%%%%%%%%%%%%%%%%%%%%%%%%%
\section*{Numerical Method of Solution}

In the present study, the flow problem is modeled by ensemble-averaged continuity, full compressible Navier-Stokes and energy equations which are valid for calorically perfect gas. Turbulence is modeled by realizable k-$\epsilon$. Enhanced wall treatment is also used in this model for better prediction of the boundary layer over the plate. \par
Fluent 14.0.0 is used to obtain the solution of governing equations. Pressure-velocity coupling was done. As we are interested in the steady state solutions,SIMPLE algorithm developed by Spalding and Patankar is used. The discretization scheme used was second-order upwind interpolation. Solution is continued till the residuals reached the convergence of $10^{-8}$ for energy, $10^{-6}$ for velocities, and $10^{-5}$ for turbulence quantities.

%%%%%%%%%%%%%%%%%%%%%%%%%%%%%%%%%%%%%%%%%%%%%%%%%%%%%%%%%%%%%%%%%%%%%%
\section*{Grid Independence Study}

Unstructured Hexahedron Meshing of the geometry is done by ICEM-CFD (commercial software). Accuracy of  flow physics is very strong function of grid quality.  To ensure that the results are independent of the grid size. Three grids are investigated with different sizes, which are Grid1 with 2144491 elements, Grid2 with 2825292 elements and Grid3 with 5220375 elements. The size of the grid is increased from the baseline grid to two finer grids. \par

Centerline adiabatic film cooling effectiveness is plotted for all three grids as shown in Fig. 7. It is noted that Grid1 is giving the same results as that of Grid2 and Grid3, there is only a small percentage of change in the results. Grid1 and Grid2 show a deviation of 0.34% while the error between the results for Grid1 and Grid3 is 0.22%. As the error is very small Grid1 is used for the further studies. 

\begin{figure}
\scalebox{1}{\includegraphics[width=3.25in]{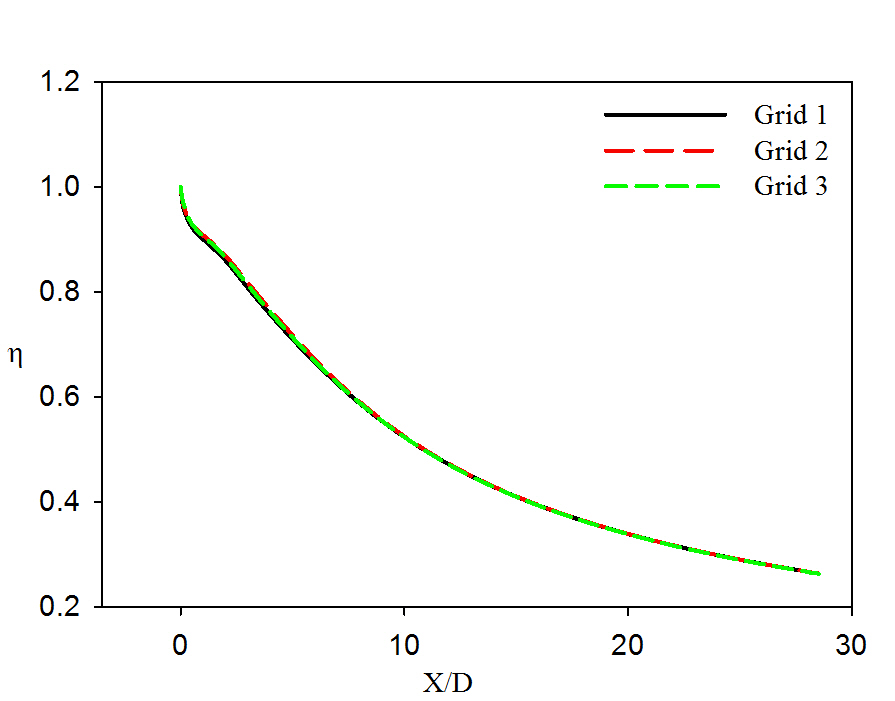}}
  %\centerline{\includegraphics{Fig6(a).eps}}% Images in 100% size
  \caption{Centerline adiabatic effectiveness for Grid1, Grid2 and Grid3}
\label{fig7}
\end{figure}

Grid sensitivity is very important to ensure the correct predictions of the laws of physics. As to achieve it, the region near the wall has finer grid for resolving the boundary layer as shown in Fig. 8. Similarly, the region of the interaction of the cold and free stream is also has finer mesh. The regions far from the flat plate can have coarse grid but the transition is kept to be smoother. Fig. 9 shows the grid around the film cooling for both normal injection and compound hole.  \par

\begin{figure}
\scalebox{1}{\includegraphics[width=3.25in]{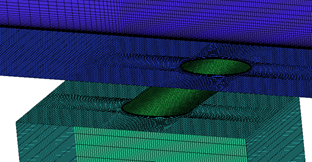}}
  %\centerline{\includegraphics{Fig6(a).eps}}% Images in 100% size
 \begin{center}
  (a)\\
  \end{center}
\label{fig8a}

\scalebox{1}{\includegraphics[width=3.25in]{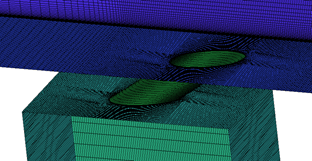}}
  %\centerline{\includegraphics{Fig6(a).eps}}% Images in 100% size
  \begin{center}
  (b)\\
  \end{center}
  \caption{Grid used in present study: (a) normal injection angle; (b) compound angle $\beta$ = $30\degree$}
\label{fig8b}
\end{figure}

\begin{figure}
\scalebox{1}{\includegraphics[width=3.25in]{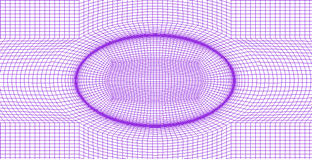}}
  %\centerline{\includegraphics{Fig6(a).eps}}% Images in 100% size
  \begin{center}
  (a)\\
  \end{center}
  %\caption{Grid near the film cooling hole: (a) of normal injection; (b) of compound angle $\beta$ = $30\degree$}
\label{fig9a}
\end{figure}

\begin{figure}
\scalebox{1}{\includegraphics[width=3.25in]{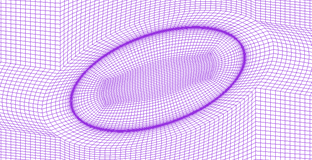}}
  %\centerline{\includegraphics{Fig6(a).eps}}% Images in 100% size
  \begin{center}
  (b)\\
  \end{center}
  \caption{Grid near the film cooling hole: (a) of normal injection; (b) of compound angle $\beta$ = $30\degree$}
\label{fig9b}
\end{figure} 

The first layer of the elements on the wall and near the hole for all grids has y+ below 1. Fig.10 shows the value of $y^+$  near the hole on the plate, this value of $y^+$ is calculated for $\alpha $= $35\degree$ , $\beta$=$30\degree$ and L/D=2. The grids used for the further parametric studies have elements varying from 2144491 to 2758446. Which depends upon the compound angle and L/D ratio. \par

\begin{figure}
\scalebox{1}{\includegraphics[width=3.25in]{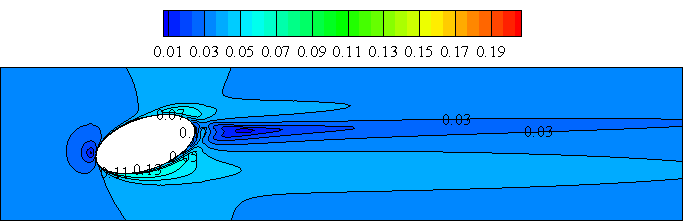}}
  %\centerline{\includegraphics{Fig6(a).eps}}% Images in 100% size
  \caption{Values of $y^+$ near the film cooling hole}
\label{fig10}
   \end{figure}

%%%%%%%%%%%%%%%%%%%%%%%%%%%%%%%%%%%%%%%%%%%%%%%%%%%%%%%%%%%%%%%%%%%%%%
\section*{Validation}

As the CFD simulation is nothing more than a mathematical predication, it is necessary to validate the model used in the simulation with some significant results. In this study the validation is done by comparing the baseline geometry results with the experimental results of Kohli and Bogard [3] and numerical results of Sangkwon and Tom S. [14] for $\alpha$=$35\degree$, L/D=2.8 at blowing ratio M=0.5. Note: for validation we have used the same hole diameter which is used by Kohli and Bogard [3] and Sangkwon and Tom S. [14], D=12.7mm. 

\begin{figure}[b]
\scalebox{1}{\includegraphics[width=3.25in]{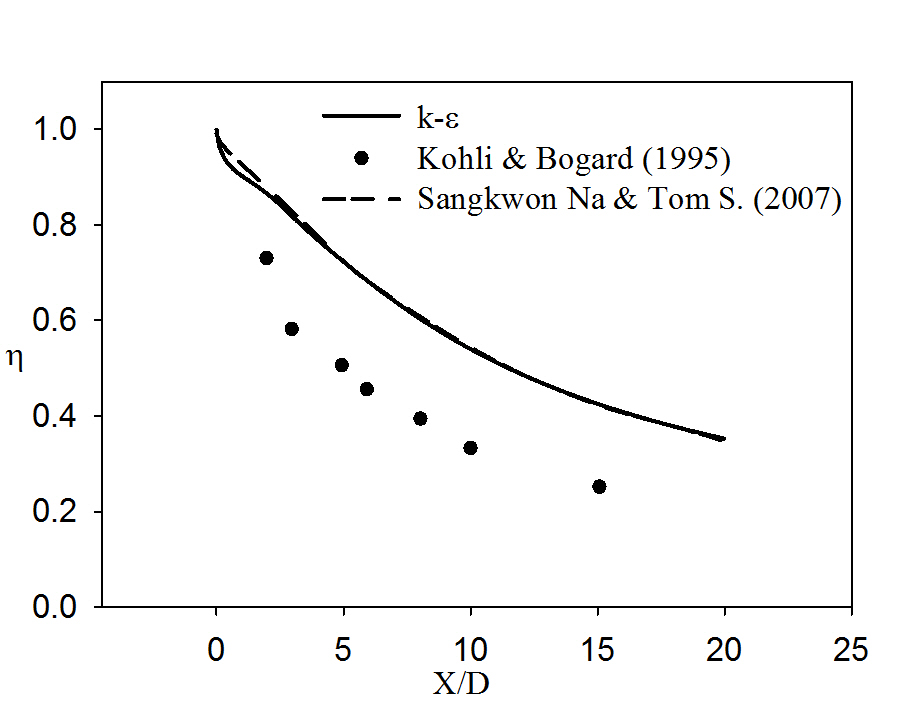}}
  %\centerline{\includegraphics{Fig6(a).eps}}% Images in 100% size
 % \caption{Centerline adiabatic effectiveness for grid1, grid2 and grid3}
 \begin{center}
  (a)\\
  \end{center}
\label{fig80}
\end{figure} 

\begin{figure}
\scalebox{1}{\includegraphics[width=3.25in]{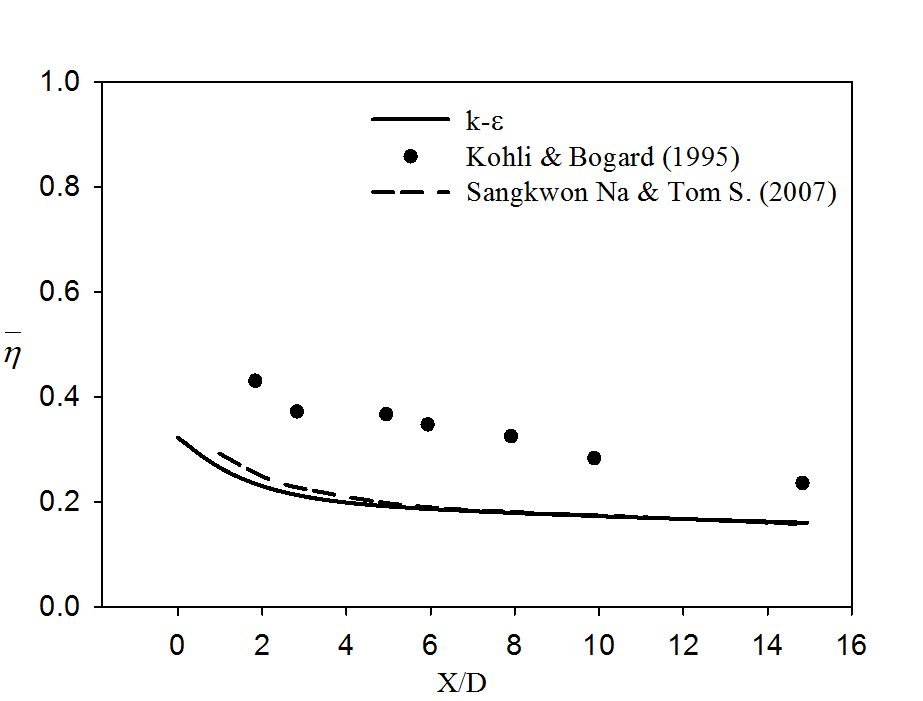}}
  %\centerline{\includegraphics{Fig6(a).eps}}% Images in 100% size
  \begin{center}
  (b)\\
  \end{center}
  \caption{Comparison of present results with the results of Kohli and bogard[3] and Sangkwon and Tom S.[14] (a) Centerline ; (b) Laterally averaged adiabatic film cooling effectiveness. }
\label{fig81}
\end{figure} 

Predicted results are shown in Fig. 11 with the results of Kohli and Bogard [3] and Sangkwon and Tom S. [14]. Which shows that the present numerical results over-predicts the experimental results of Kohli and Bogard [3] but in good agreement with Sangkwon and Tom S. [14] as shown in Fig. 11(a) for centerline adiabatic film cooling effectiveness in the downstream. However present numerical results under-predicts the experimental results of Kohli and Bogard [3] but in good agreement with Sangkwon and Tom S. [14] as shown in Fig. 11(b) for laterally averaged adiabatic film cooling effectiveness. It means in numerical model lateral spread of the coolant is less while longitudinal spread is more than the experiments. Although presented numerical results are deviated from experimental results yet numerical results are predicting same as Sangkwon and Tom S. [14].  \par

%%%%%%%%%%%%%%%%%%%%%%%%%%%%%%%%%%%%%%%%%%%%%%%%%%%%%%%%%%%%%%%%%%%%%%
\section*{Results and Discussions}

Detailed parametric study was done to analyze the effects of compound angle and L/D ratio. Parameters used to study these effects are tabulated in Table [1].

\begin{table}
\caption{Summary of simulations performed}
\begin{center}
\label{table1}
\begin{tabular}{c c c c c c }
\hline
\hline
Cases              & $\alpha$ & $\beta $ & L/D  & M  & DR  \\
\hline
1& $35\degree$ & $0\degree$ & 1 & 0.5 & 1.6  \\
2& $35\degree$ & $0\degree$ & 2 & 0.5 & 1.6   \\
3& $35\degree$ & $0\degree$ & 3 & 0.5 & 1.6    \\
4& $35\degree$ & $0\degree$ & 4 & 0.5 & 1.6    \\
5& $35\degree$ & $30\degree$ & 1 & 0.5 & 1.6    \\
6& $35\degree$ & $30\degree$ & 2 & 0.5 & 1.6    \\
7& $35\degree$ & $30\degree$ & 3 & 0.5 & 1.6    \\
8& $35\degree$ & $30\degree$ & 4 & 0.5 & 1.6    \\
9& $35\degree$ & $45\degree$ & 1 & 0.5 & 1.6   \\
10& $35\degree$ & $45\degree$ & 2 & 0.5 & 1.6   \\
11& $35\degree$ & $45\degree$ & 3 & 0.5 & 1.6   \\
12& $35\degree$ & $45\degree$ & 4 & 0.5 & 1.6   \\
13& $35\degree$ & $60\degree$ & 1 & 0.5 & 1.6   \\
14& $35\degree$ & $60\degree$ & 2 & 0.5 & 1.6   \\
15& $35\degree$ & $60\degree$ & 3 & 0.5 & 1.6   \\
16& $35\degree$ & $60\degree$ & 4 & 0.5 & 1.6   \\

\hline
\hline
\end{tabular}
\end{center}
\end{table}

The results obtained from the parametric study shows very significant effects on the flow pattern and adiabatic film cooling effectiveness. 

\begin{figure}[h!]
\scalebox{1}{\includegraphics[width=3.25in]{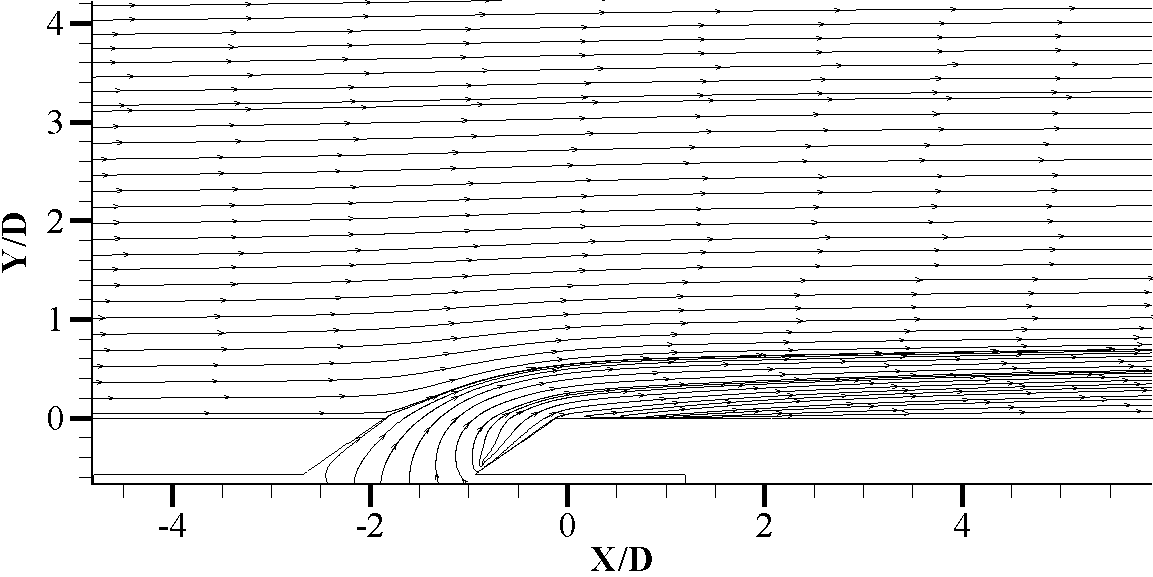}}
  %\centerline{\includegraphics{Fig6(a).eps}}% Images in 100% size
  \begin{center}
  (a)\\
  \end{center}
  %\caption{Grid near the film cooling hole: (a) of normal injection; (b) of compound angle $\beta$=300}
\label{fig12a}

\scalebox{1}{\includegraphics[width=3.25in]{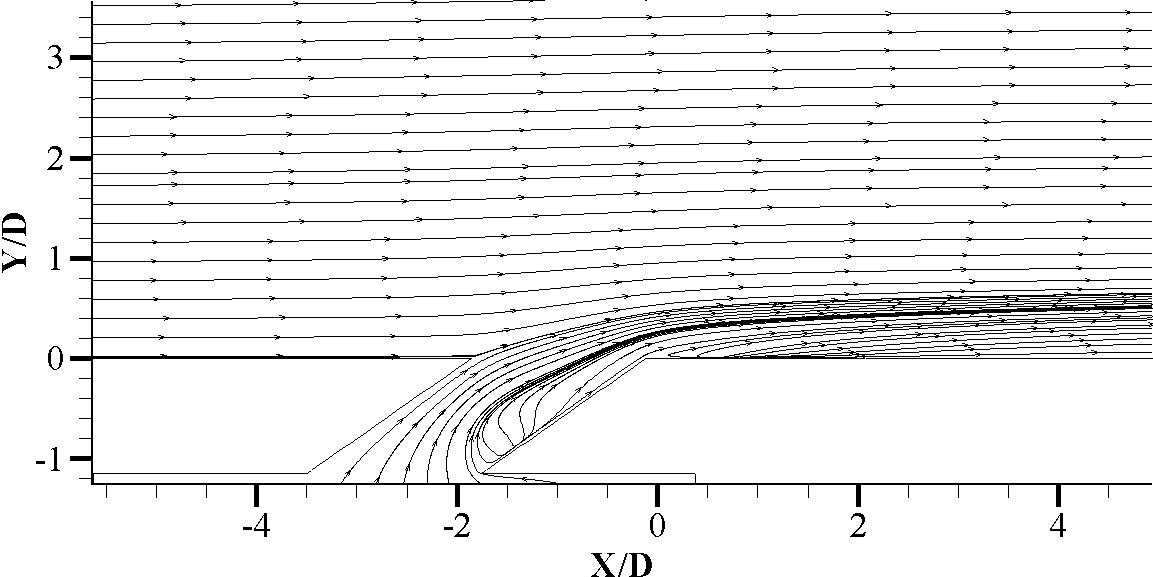}}
  %\centerline{\includegraphics{Fig6(a).eps}}% Images in 100% size
  \begin{center}
  (b)\\
  \end{center}
  %\caption{Grid near the film cooling hole: (a) of normal injection; (b) of compound angle $\beta$=300}
\label{fig12b}

\scalebox{1}{\includegraphics[width=3.25in]{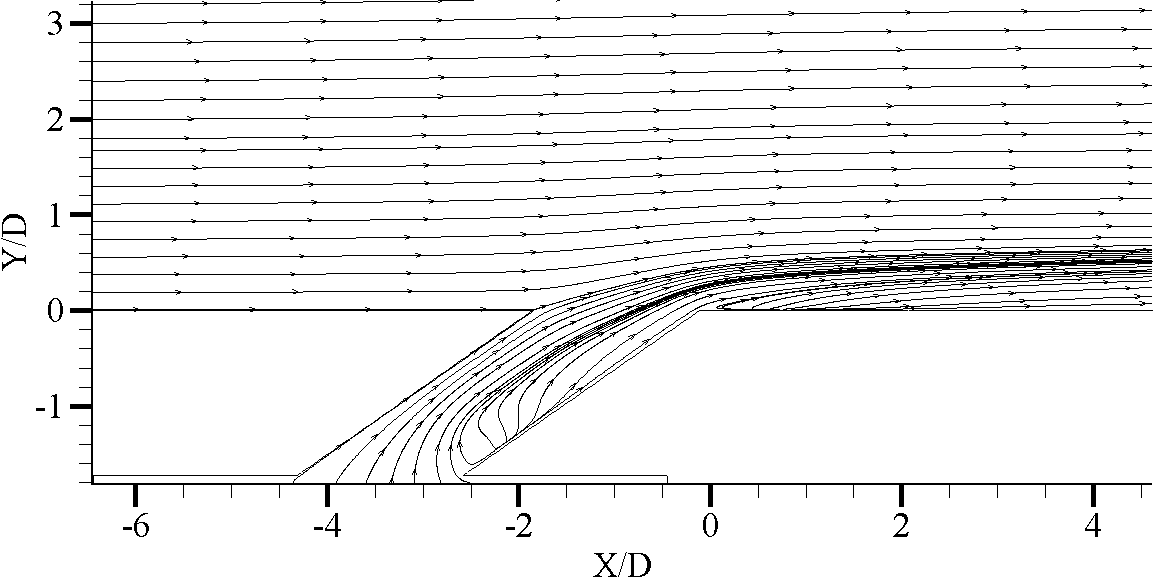}}
  %\centerline{\includegraphics{Fig6(a).eps}}% Images in 100% size
  \begin{center}
  (c)\\
  \end{center}
  %\caption{Grid near the film cooling hole: (a) of normal injection; (b) of compound angle $\beta$=300}
\label{fig12c}

\scalebox{1}{\includegraphics[width=3.25in]{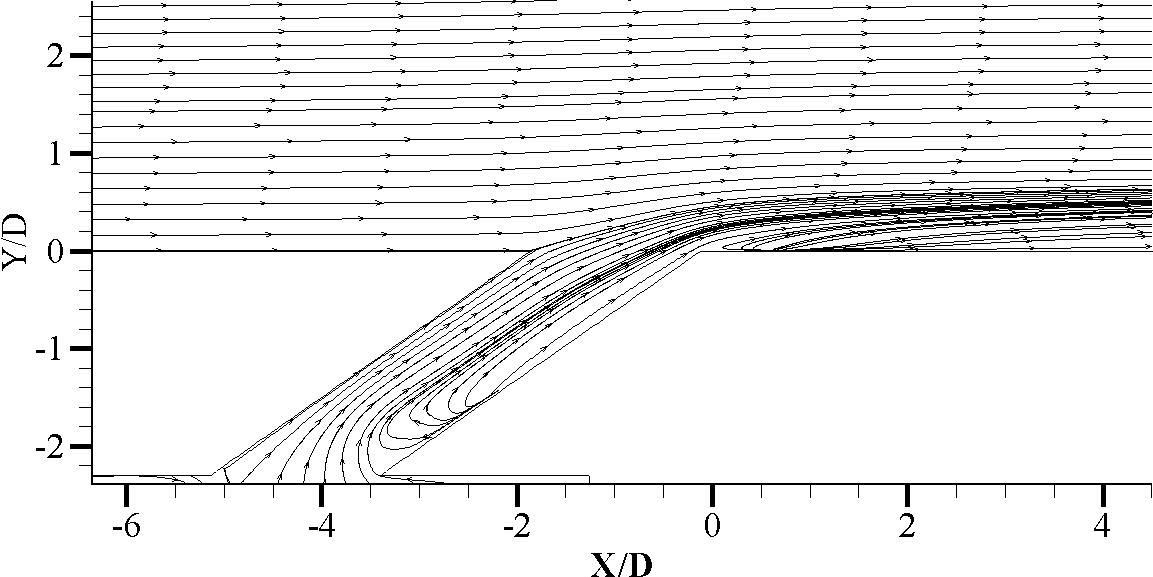}}
  %\centerline{\includegraphics{Fig6(a).eps}}% Images in 100% size
  \begin{center}
  (d)\\
  \end{center}
  \caption{Streamlines at Z/D=0 plane for $\alpha$ =$35\degree$ with: (a) L/D=1; (b) L/D=2; (c) L/D=3; (d) L/D=4 }
\label{fig12d}
\end{figure} 

{\bf Effect of L/D ratio on flow pattern.}
For small L/D ratio, flow comes out of hole abruptly as it does not get enough length to settle down from the disturbance caused by the flow between the plenum and hole. Same can be seen in Fig. 12(a) and (b) for L/D=1 and 2. As L/D ratio increases flow gets time to settle in the tube. This makes the flow to be more towards the leading edge of the hole then trailing edge as shown in Fig. 12(c)-(d) for L/D=3 and 4, it is settled before it exits the hole. \par
 For large L/D ratios, as the flow is towards leading edge which results more effective penetration of the cold stream in downstream direction and give more effective spread up to larger X/Ds in downstream. \par

\begin{figure}[h!]
\scalebox{1}{\includegraphics[ trim= 0 0 2mm 0, clip=true, width=3.25in]{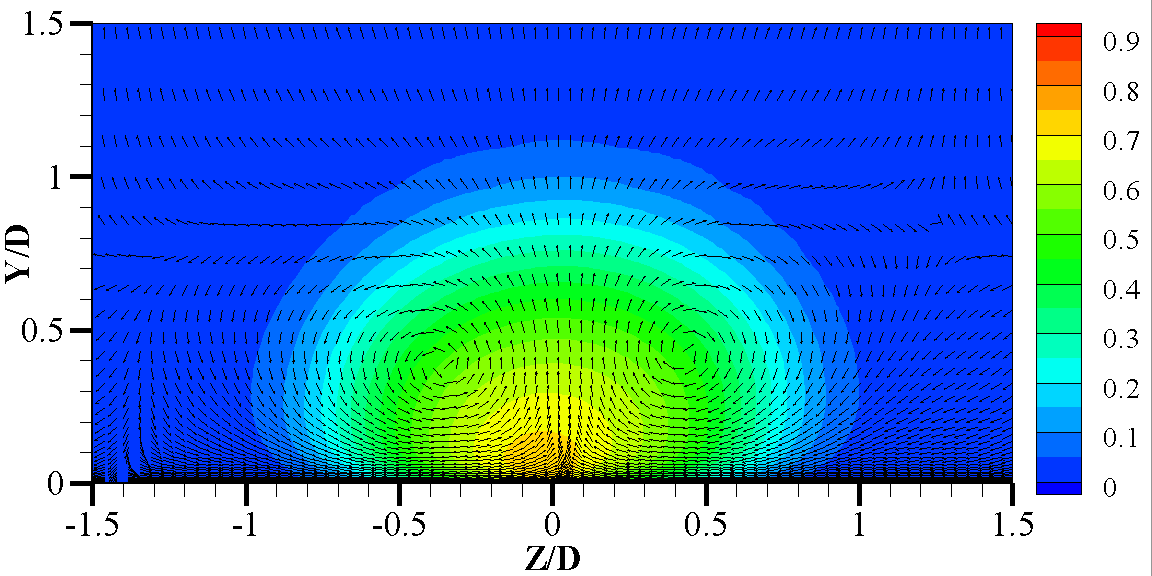}}
  %\centerline{\includegraphics{Fig6(a).eps}}% Images in 100% size
  \begin{center}
  (a)\\
  \end{center}
  %\caption{Grid near the film cooling hole: (a) of normal injection; (b) of compound angle $\beta$=300}
\label{fig13a}

\scalebox{1}{\includegraphics[trim= 0 0 2mm 0, clip=true,width=3.25in]{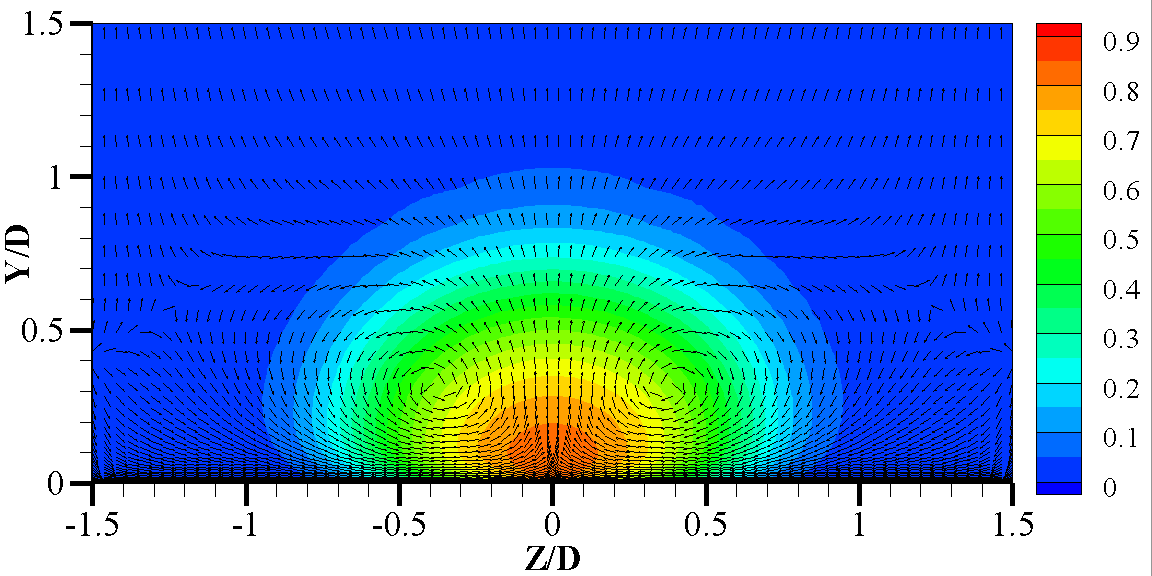}}
  %\centerline{\includegraphics{Fig6(a).eps}}% Images in 100% size
  \begin{center}
  (b)\\
  \end{center}
  %\caption{Grid near the film cooling hole: (a) of normal injection; (b) of compound angle $\beta$=300}
\label{fig13b}

\scalebox{1}{\includegraphics[trim= 0 0 2mm 0, clip=true,width=3.25in]{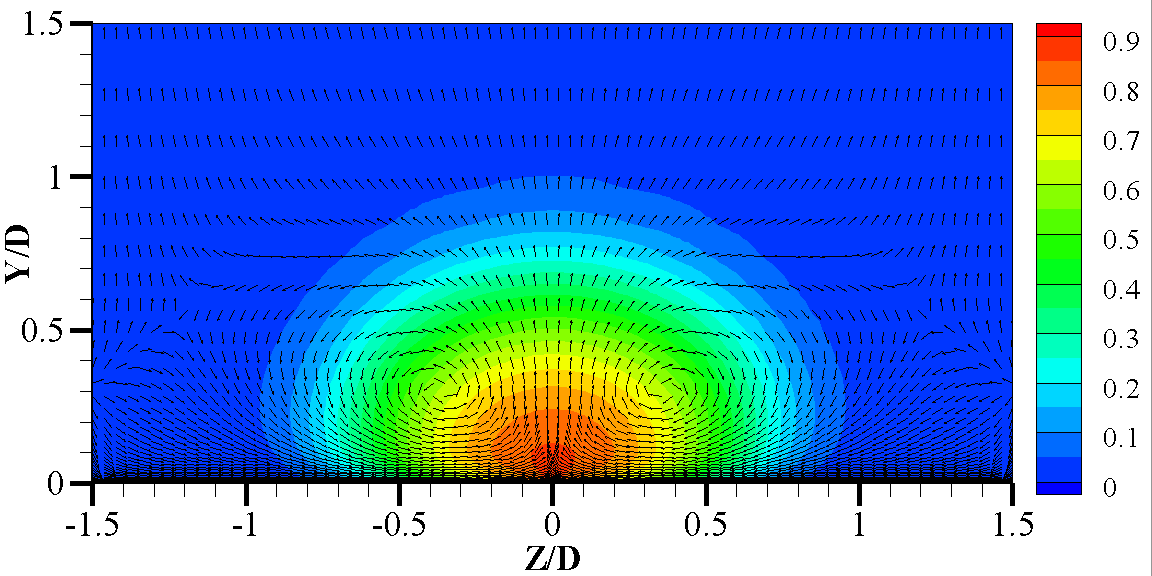}}
  %\centerline{\includegraphics{Fig6(a).eps}}% Images in 100% size
  \begin{center}
  (c)\\
  \end{center}
  %\caption{Grid near the film cooling hole: (a) of normal injection; (b) of compound angle $\beta$=300}
\label{fig13c}

\scalebox{1}{\includegraphics[trim= 0 0 2mm 0, clip=true,width=3.25in]{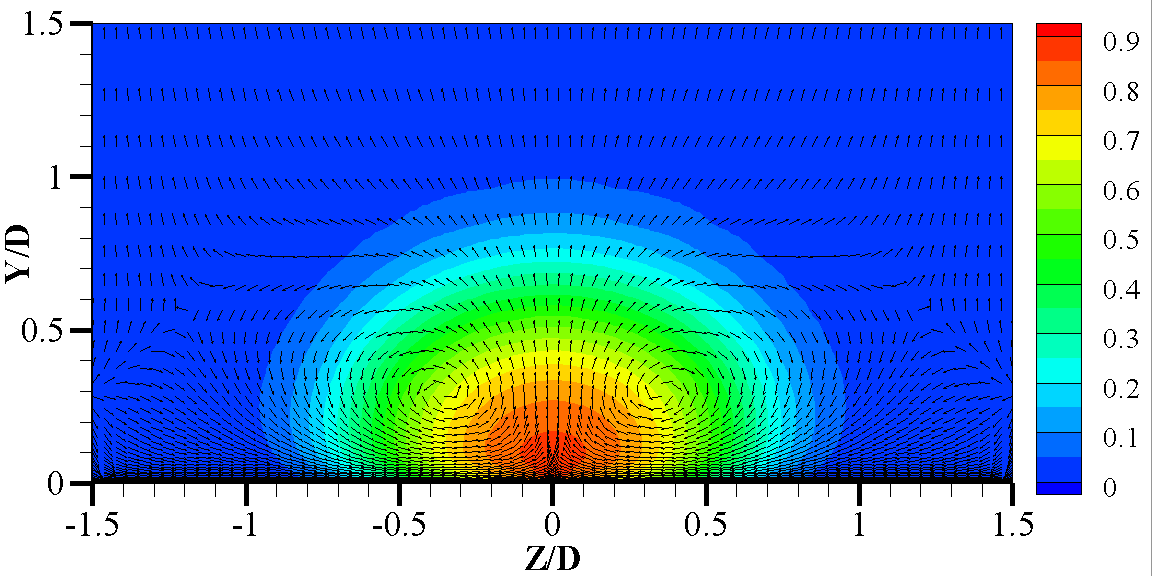}}
  %\centerline{\includegraphics{Fig6(a).eps}}% Images in 100% size
  \begin{center}
  (d)\\
  \end{center}
  \caption{Contours of normalized temperature at X/D=3 plane for $\beta$ =$0\degree$ with: (a) L/D=1; (b) L/D=2; (c) L/D=3; (d) L/D=4 }
 % \caption{Contours of normalized temperature at X/D=3 plane for }%$\beta$ =$0\degree$ with: (a) L/D=1; (b) L/D=2; (c) L/D=3; (d) L/D=4.}
\label{fig13d}
\end{figure}  

Increase in L/D ratio also ensure the less penetration of cold stream in y direction. As it can be seen in the contours shown in Fig. 13-16, for all compound holes. For Compound angle $\beta$ =$0\degree$ and $\beta$ =$30\degree$ the reduction of penetration in y direction is very small as seen in Fig. 13-14. As compound angle increases to $\beta$ =$45\degree$ and $\beta$ =$60\degree$ as shown in Fig. 15-16, the penetration in y direction is reduced from Y/D=1 and 1.25 for L/D=1 to Y/D=0.95 and 0.85 for L/D=4 respectively.

\begin{figure}[h!]
\scalebox{1}{\includegraphics[trim= 0 0 2mm 0, clip=true,width=3.25in]{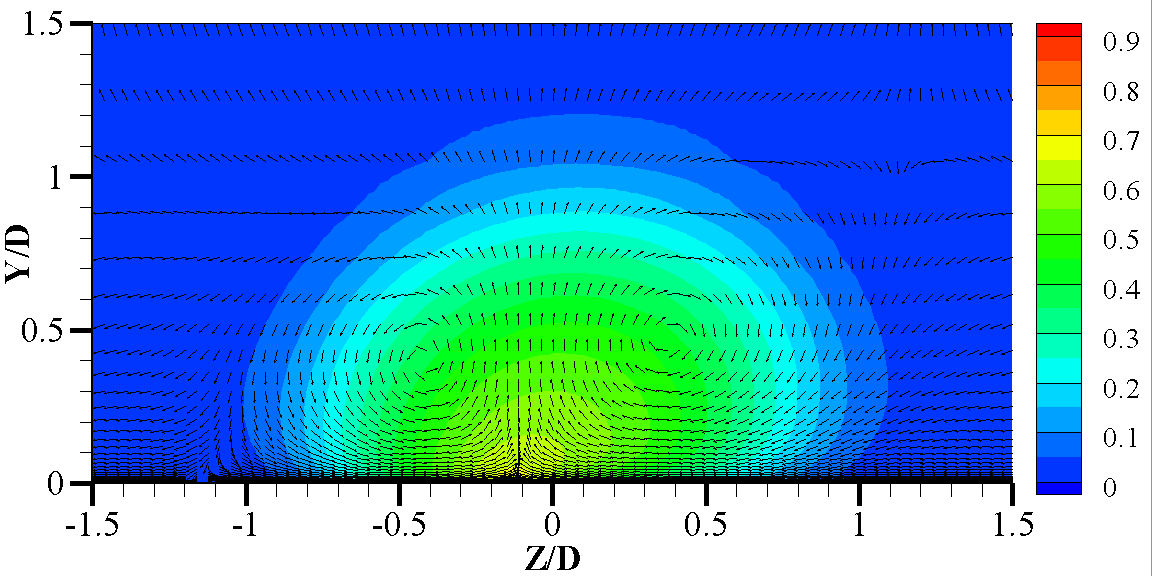}}
  %\centerline{\includegraphics{Fig6(a).eps}}% Images in 100% size
  \begin{center}
  (a)\\
  \end{center}
  %\caption{Grid near the film cooling hole: (a) of normal injection; (b) of compound angle $\beta$=300}
\label{fig14a}

\scalebox{1}{\includegraphics[trim= 0 0 2mm 0, clip=true,width=3.25in]{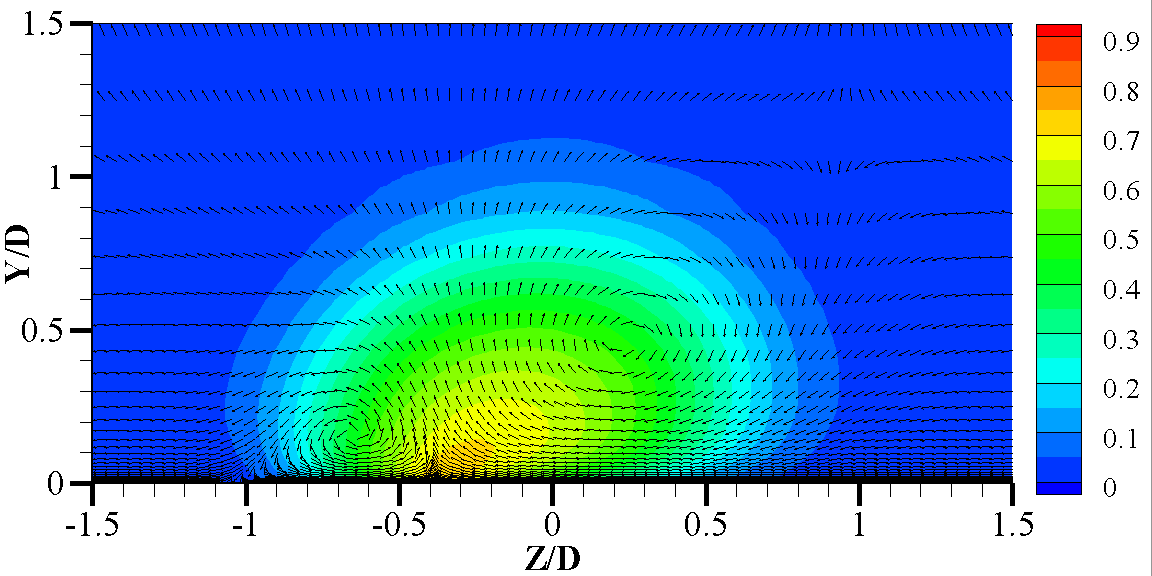}}
  %\centerline{\includegraphics{Fig6(a).eps}}% Images in 100% size
  \begin{center}
  (b)\\
  \end{center}
  %\caption{Grid near the film cooling hole: (a) of normal injection; (b) of compound angle $\beta$=300}
\label{fig14b}

\scalebox{1}{\includegraphics[trim= 0 0 2mm 0, clip=true,width=3.25in]{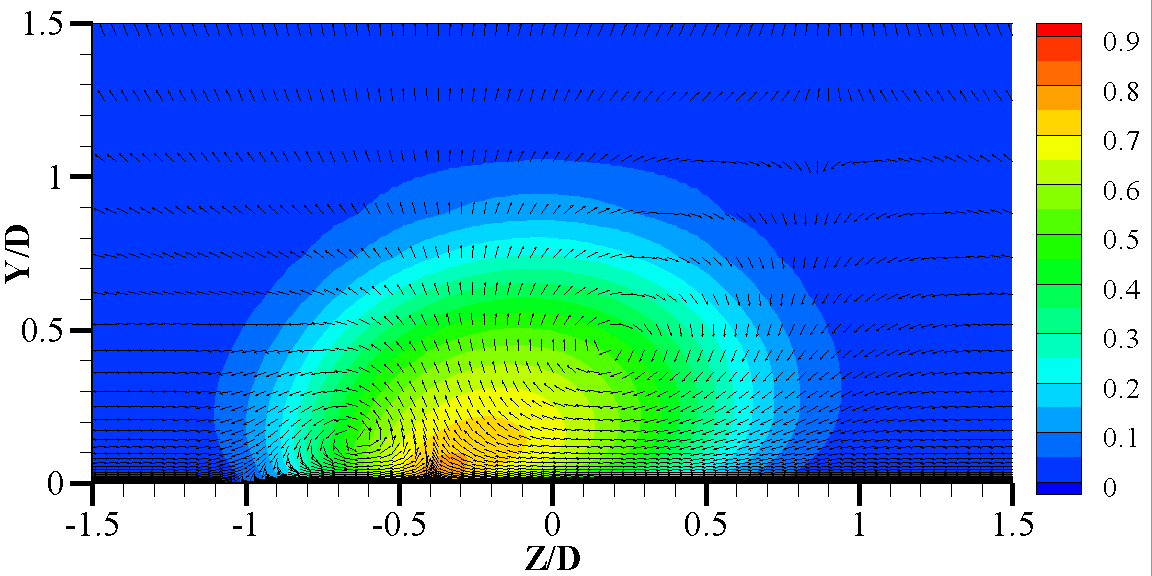}}
  %\centerline{\includegraphics{Fig6(a).eps}}% Images in 100% size
  \begin{center}
  (c)\\
  \end{center}
  %\caption{Grid near the film cooling hole: (a) of normal injection; (b) of compound angle $\beta$=300}
\label{fig14c}

\scalebox{1}{\includegraphics[trim= 0 0 2mm 0, clip=true,width=3.25in]{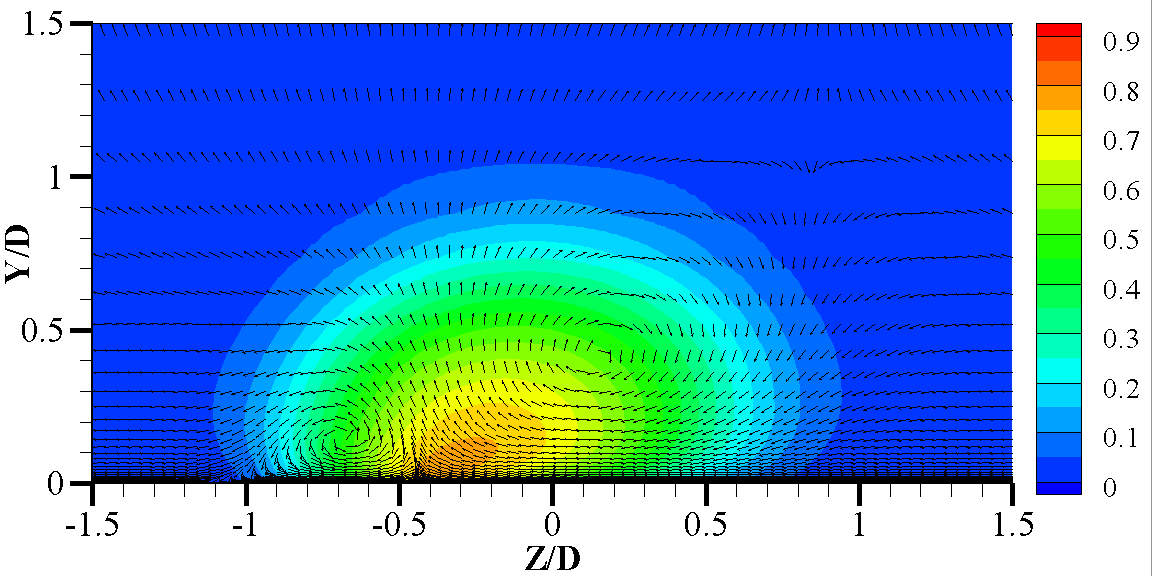}}
  %\centerline{\includegraphics{Fig6(a).eps}}% Images in 100% size
  \begin{center}
  (d)\\
  \end{center}
  \caption{Contours of normalized temperature at X/D=3 plane for $\beta$ =$30\degree$ with: (a) L/D=1; (b) L/D=2; (c) L/D=3; (d) L/D=4 }
  %\caption{Contours of normalized temperature at X/D=3 plane for}% $\beta = 30\degree$ with: (a) L/D=1; (b) L/D=2; (c) L/D=3; (d) L/D=4.}
\label{fig14d}
\end{figure} 

It can also be seen from the obtained results that the spread of the cold stream shifts from  right to left in lateral direction as observed in the direction of downstream, this happens due to increase in L/D ratio, which settle the flow according to the compounded hole as shown in Fig. 18-20. This does not happen for $\beta$ =$0\degree$  as it is the normal injection hole where the cold stream is only vectored in longitudinal direction shown in Fig. 17. Unlikely in the case of $\beta$ =$30\degree$ , $\beta$ =$45\degree$ and $\beta$ =$60\degree$  as the compound angle increases the vectoring of cold stream in lateral direction  increases. This effect will be more if we also increase  L/D ratio for the given compound angle, which makes the flow to be more focused in lateral direction. Thus, increase in compound angle with increase in L/D ratio increases the lateral shifting of the cold stream which can easily be seen in Fig 18-20.     

\begin{figure}[h!]
\scalebox{1}{\includegraphics[trim= 0 0 2mm 0, clip=true,width=3.25in]{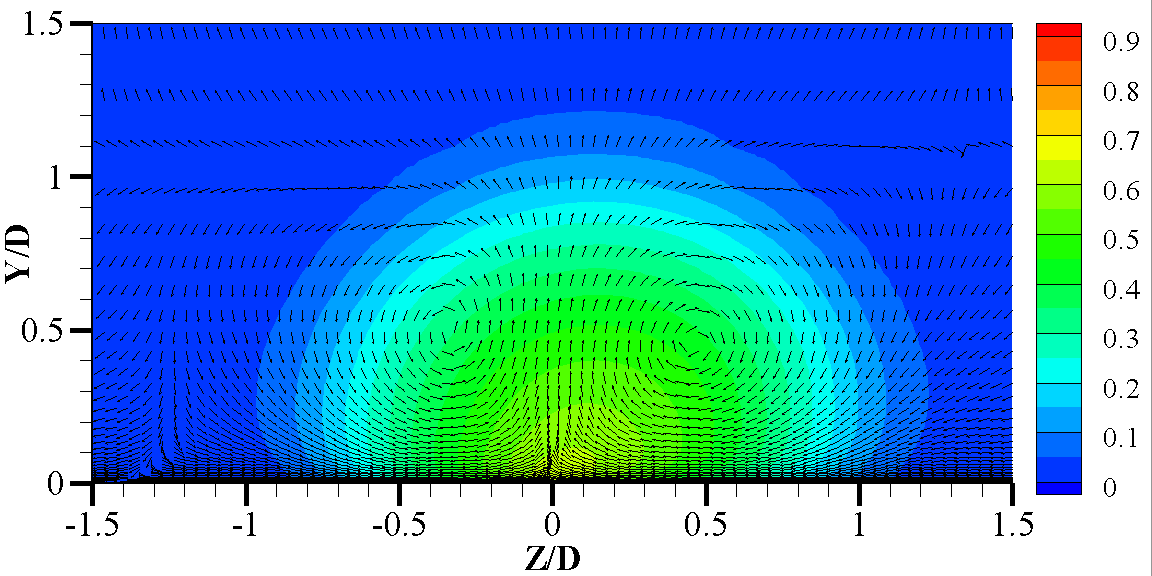}}
  %\centerline{\includegraphics{Fig6(a).eps}}% Images in 100% size
  \begin{center}
  (a)\\
  \end{center}
  %\caption{Grid near the film cooling hole: (a) of normal injection; (b) of compound angle $\beta$=300}
\label{fig15a}

\scalebox{1}{\includegraphics[trim= 0 0 2mm 0, clip=true,width=3.25in]{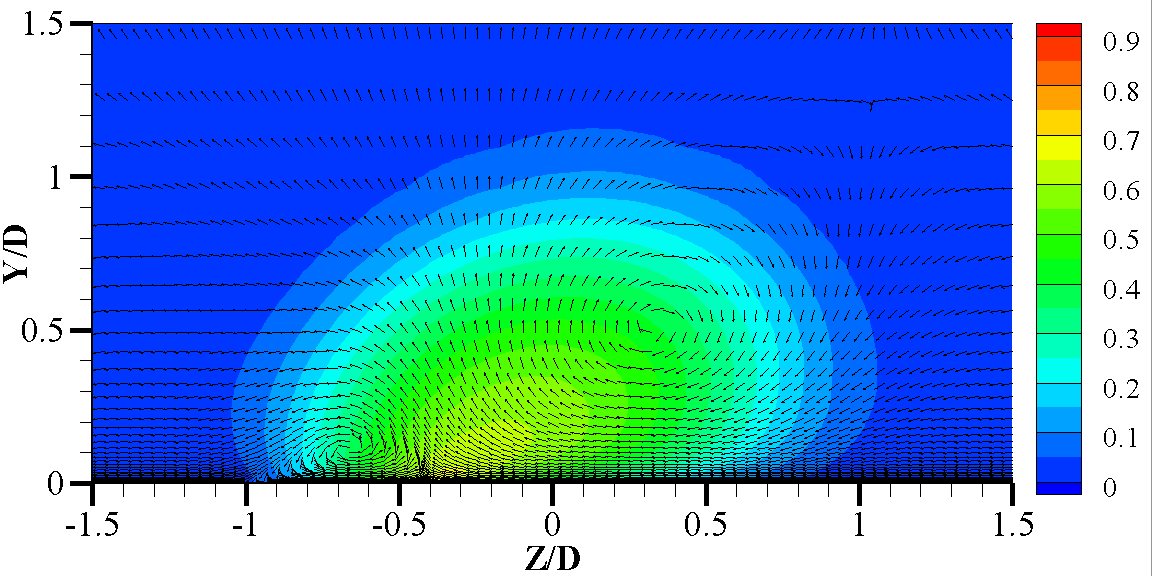}}
  %\centerline{\includegraphics{Fig6(a).eps}}% Images in 100% size
  \begin{center}
  (b)\\
  \end{center}
  %\caption{Grid near the film cooling hole: (a) of normal injection; (b) of compound angle $\beta$=300}
\label{fig15b}

\scalebox{1}{\includegraphics[trim= 0 0 2mm 0, clip=true,width=3.25in]{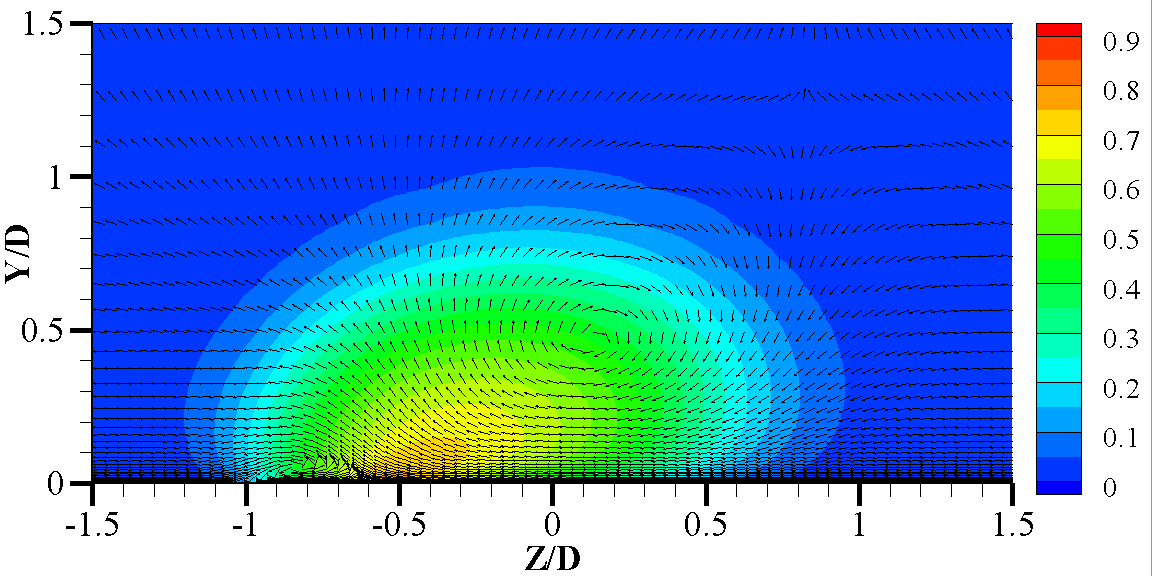}}
  %\centerline{\includegraphics{Fig6(a).eps}}% Images in 100% size
  \begin{center}
  (c)\\
  \end{center}
  %\caption{Grid near the film cooling hole: (a) of normal injection; (b) of compound angle $\beta$=300}
\label{fig15c}

\scalebox{1}{\includegraphics[trim= 0 0 2mm 0, clip=true,width=3.25in]{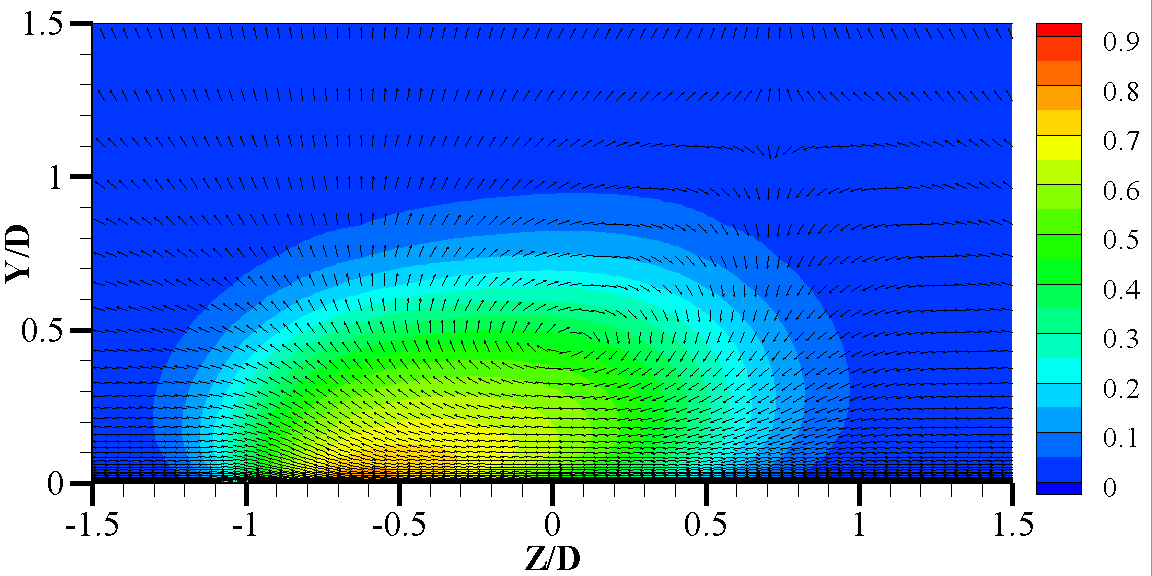}}
  %\centerline{\includegraphics{Fig6(a).eps}}% Images in 100% size
  \begin{center}
  (d)\\
  \end{center}
  \caption{Contours of normalized temperature at X/D=3 plane for  $\beta$ =$45\degree$ with: (a) L/D=1; (b) L/D=2; (c) L/D=3; (d) L/D=4 }
  %\caption{Contours of normalized temperature at X/D=3 plane for }%$\beta=45\degree$ with: (a) L/D=1; (b) L/D=2; (c) L/D=3; (d) L/D=4.}
\label{fig15d}
\end{figure}

\begin{figure}[h!]
\scalebox{1}{\includegraphics[trim= 0 0 2mm 0, clip=true,width=3.25in]{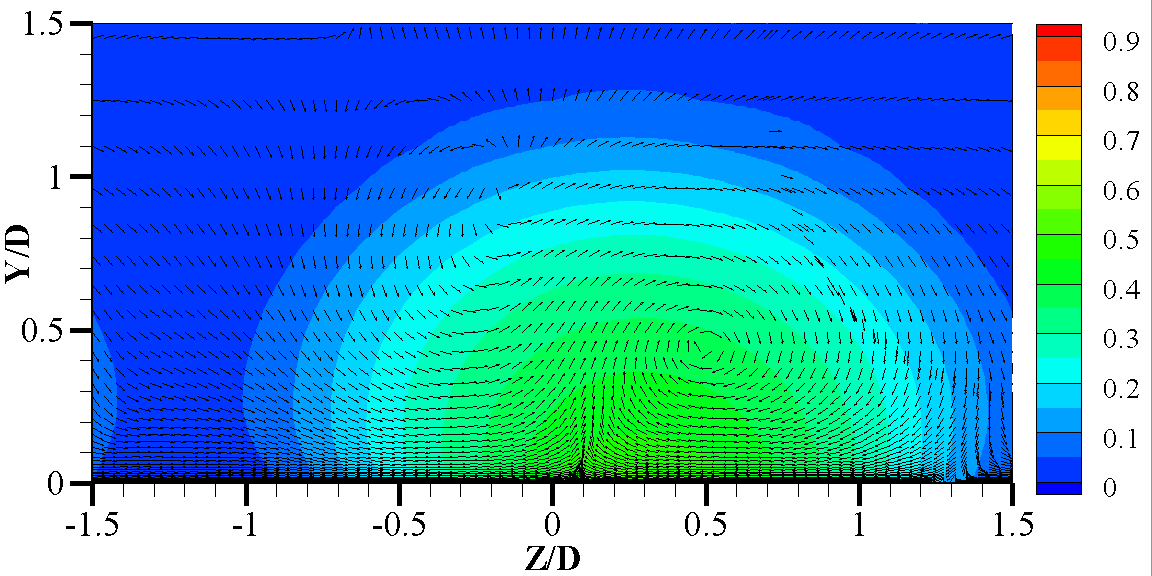}}
  %\centerline{\includegraphics{Fig6(a).eps}}% Images in 100% size
  \begin{center}
  (a)\\
  \end{center}
  %\caption{Grid near the film cooling hole: (a) of normal injection; (b) of compound angle $\beta$=300}
\label{fig16a}

\scalebox{1}{\includegraphics[trim= 0 0 2mm 0, clip=true,width=3.25in]{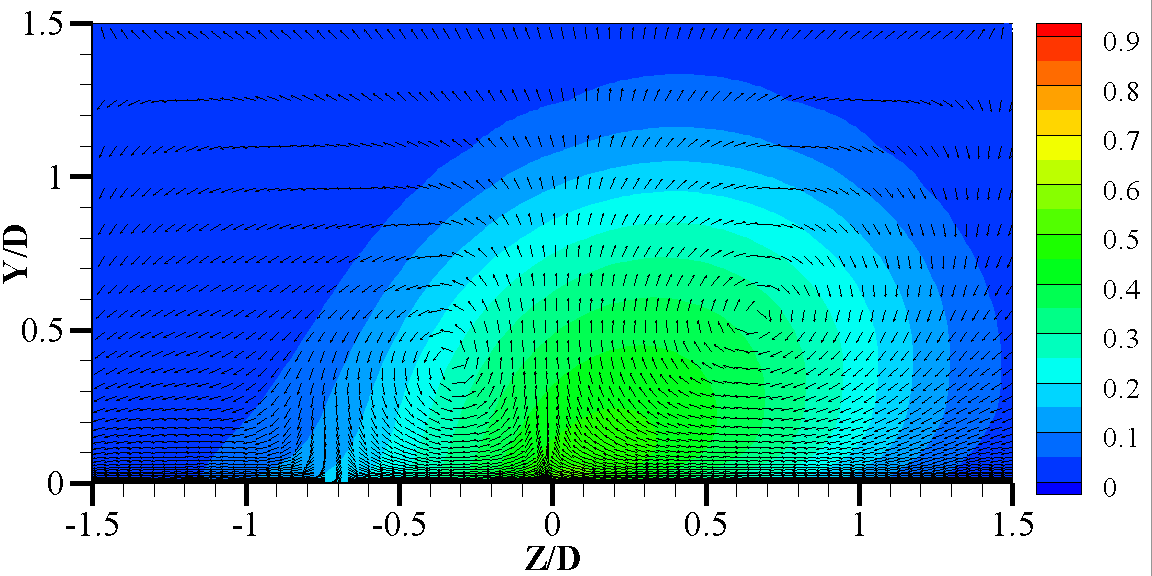}}
  %\centerline{\includegraphics{Fig6(a).eps}}% Images in 100% size
  \begin{center}
  (b)\\
  \end{center}
  %\caption{Grid near the film cooling hole: (a) of normal injection; (b) of compound angle $\beta$=300}
\label{fig16b}

\scalebox{1}{\includegraphics[trim= 0 0 2mm 0, clip=true,width=3.25in]{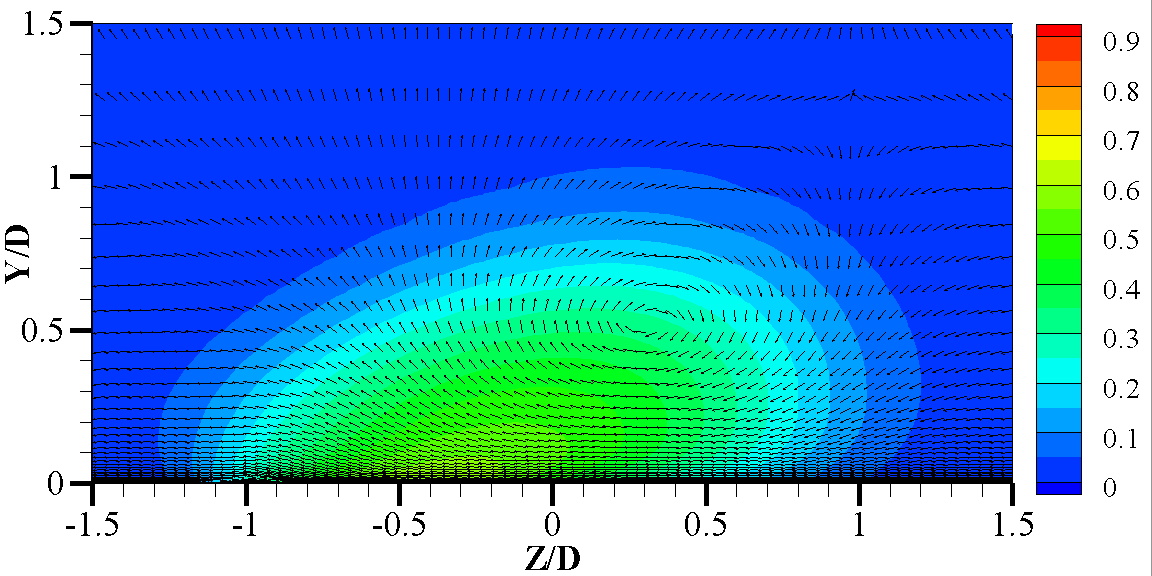}}
  %\centerline{\includegraphics{Fig6(a).eps}}% Images in 100% size
  \begin{center}
  (c)\\
  \end{center}
  %\caption{Grid near the film cooling hole: (a) of normal injection; (b) of compound angle $\beta$=300}
\label{fig16c}

\scalebox{1}{\includegraphics[trim= 0 0 2mm 0, clip=true,width=3.25in]{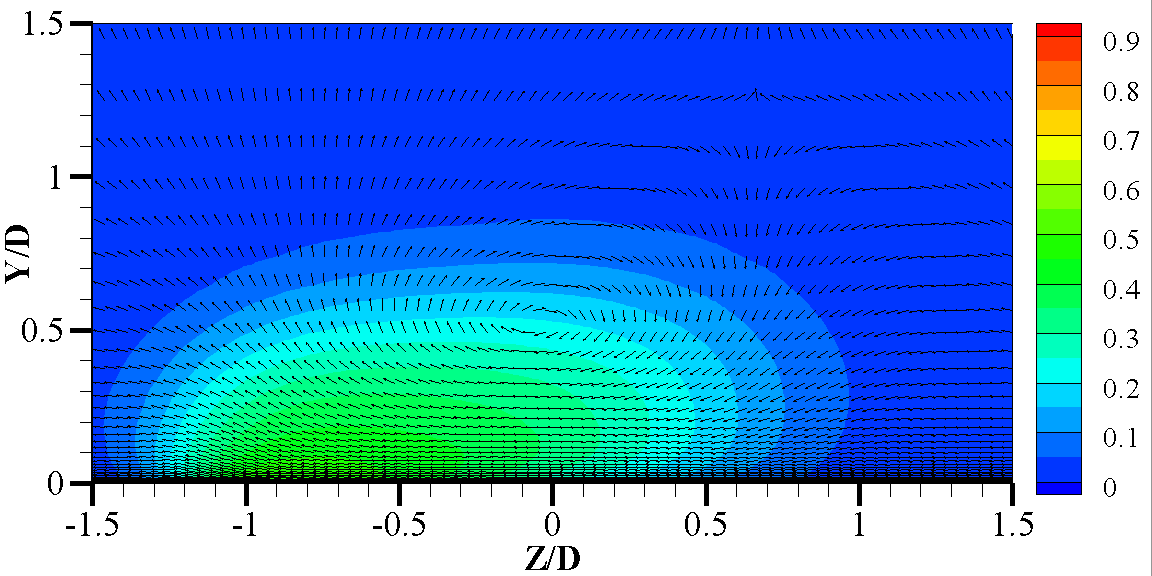}}
  %\centerline{\includegraphics{Fig6(a).eps}}% Images in 100% size
  \begin{center}
  (d)\\
  \end{center}
  \caption{Contours of normalized temperature at X/D=3 plane for $\beta$ =$60\degree$ with: (a) L/D=1; (b) L/D=2; (c) L/D=3; (d) L/D=4 }
 % \caption{Contours of normalized temperature at X/D=3 plane for }%$\beta$ =$60\degree$ with: (a) L/D=1; (b) L/D=2; (c) L/D=3; (d) L/D=4.}
\label{fig16d}
\end{figure} 

{ \bf Effect of Compound Angle on Flow. }
As it has been seen in several studies [3],[5-7] that for normal injection holes when cold and hot stream interacts there is formation of two kidney vortices Fig. 13(a)-(b). But due to compound angle in the present study, we have analyzed that the vortex in the direction of lateral vector of cold stream, which can be seen in left side with respect to Z/D=0, Fig. 14-16, starts diminishing due to effect of laterally vectored flow of cold stream. The reason behind that is initially with no compound angle the velocity of the cold flow is in the longitudinal direction which induces high velocity gradient(hot-cold velocity layer) flow causes vortex but with compound angle the net velocity of cold flow in the longitudinal direction decrease which induces less velocity gradient hence weak vortex. Secondly  this vortex is carried in lateral direction along cold flow and due to that it starts dissipating from the shear of longitudinal hot flow. For L/D=1, the reduction of the one vortex is not come into picture due the unsettled flow in hole as shown in Fig. 14(a), 15(a), 16(a). But as L/D ratio increases the effect of the compound angle increases with strong effect on kidney vortices. As for L/D=2,3 and 4 for all compound angles the vortex in left side with respect to Z/D=0 loses its intensity which can be seen in Fig. 14-16. And for L/D=4 it almost vanishes leaving only one vortex. This reduction from two to one vortex decreases the mixing of cold and hot stream. \par

\begin{figure}
\scalebox{1}{\includegraphics[trim= 2mm 0 2mm 0, clip=true,width=3.25in]{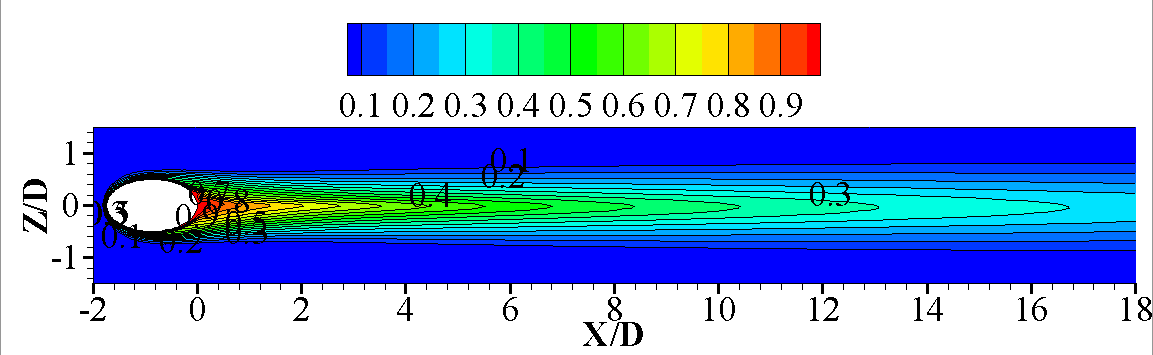}}
  %\centerline{\includegraphics{Fig6(a).eps}}% Images in 100% size
  \begin{center}
  (a)\\
  \end{center}
  %\caption{Grid near the film cooling hole: (a) of normal injection; (b) of compound angle $\beta$=300}
\label{fig17a}

\scalebox{1}{\includegraphics[trim= 2mm 0 2mm 0, clip=true,width=3.25in]{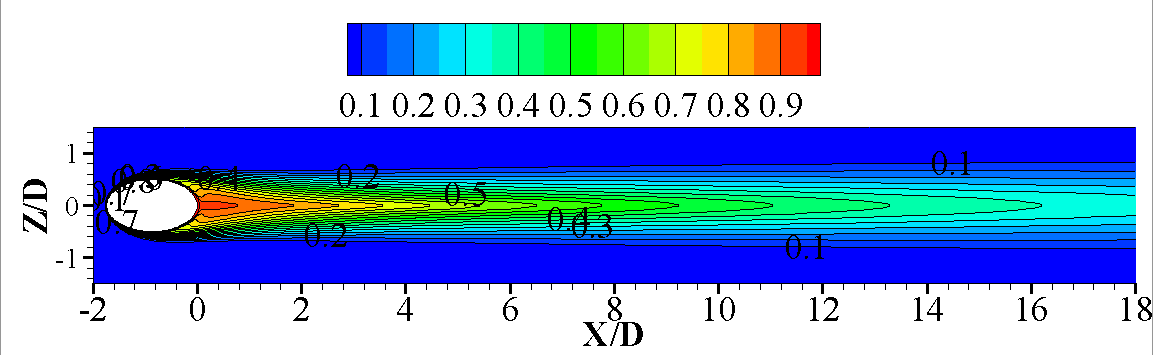}}
  %\centerline{\includegraphics{Fig6(a).eps}}% Images in 100% size
  \begin{center}
  (b)\\
  \end{center}
  %\caption{Grid near the film cooling hole: (a) of normal injection; (b) of compound angle $\beta$=300}
\label{fig17b}

\scalebox{1}{\includegraphics[trim= 2mm 0 2mm 0, clip=true,width=3.25in]{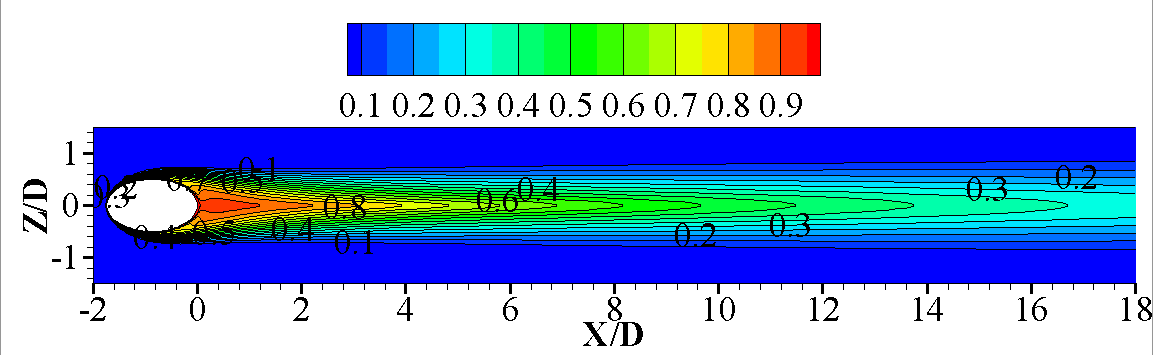}}
  %\centerline{\includegraphics{Fig6(a).eps}}% Images in 100% size
  \begin{center}
  (c)\\
  \end{center}
  %\caption{Grid near the film cooling hole: (a) of normal injection; (b) of compound angle $\beta$=300}
\label{fig17c}

\scalebox{1}{\includegraphics[trim= 2mm 0 2mm 0, clip=true,width=3.25in]{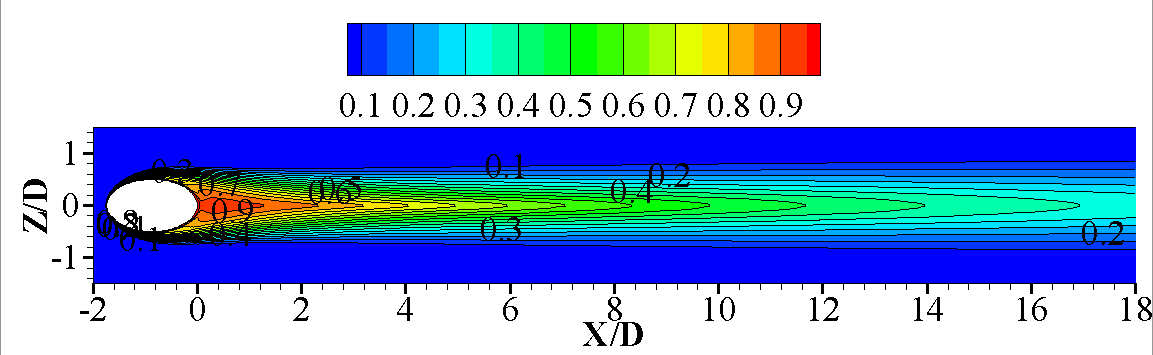}}
  %\centerline{\includegraphics{Fig6(a).eps}}% Images in 100% size
  \begin{center}
  (d)\\
  \end{center}
  \caption{Contours of adiabatic film cooling effectiveness over the plate $\beta$ =$0\degree$ with: (a) L/D=1; (b) L/D=2; (c) L/D=3; (d) L/D=4 }
\label{fig17d}
\end{figure} 

Mixing of cold flow with hot flow decreases because of dissipation of left kidney vortex and weak right kidney vortex with compound angle. Due to less mixing the cold flow easily spread to higher X/D. As we increase the L/D ratio, the effect of compound angle increase which gives further increment to spread for higher X/D. But for very higher compound angles specially in case of $\beta$=$60\degree$ with L/D=4 the intensity of the right kidney vortex is high  leading to more mixing of the cold and hot flow thus reduces the spread in longitudinal direction, which can be seen in Fig. 20(d).

\begin{figure}
\scalebox{1}{\includegraphics[trim= 2mm 0 2mm 0, clip=true,width=3.25in]{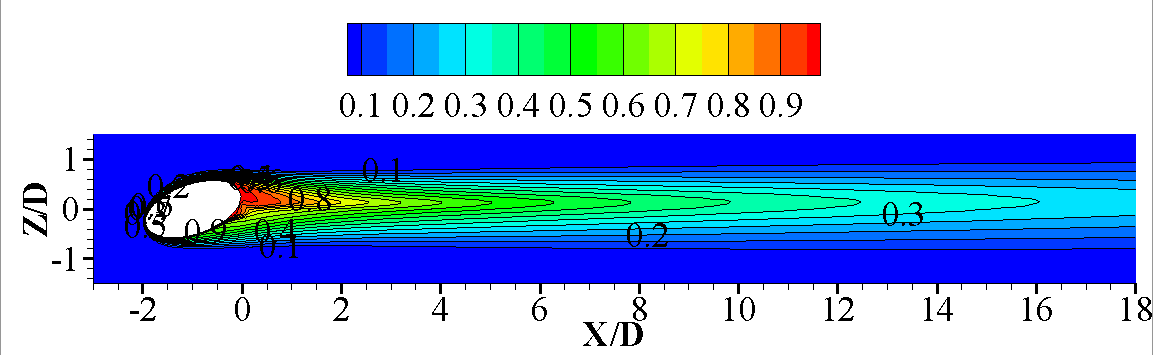}}
  %\centerline{\includegraphics{Fig6(a).eps}}% Images in 100% size
  \begin{center}
  (a)\\
  \end{center}
  %\caption{Grid near the film cooling hole: (a) of normal injection; (b) of compound angle $\beta$=300}
\label{fig18a}

\scalebox{1}{\includegraphics[trim= 2mm 0 2mm 0, clip=true,width=3.25in]{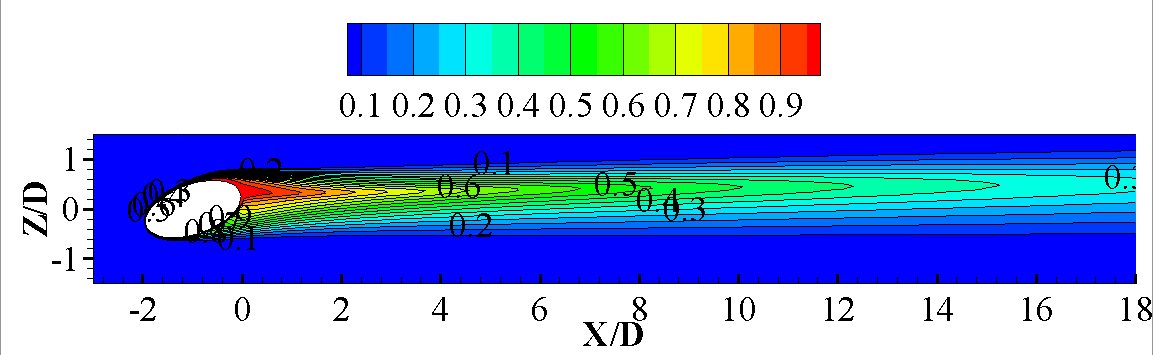}}
  %\centerline{\includegraphics{Fig6(a).eps}}% Images in 100% size
  \begin{center}
  (b)\\
  \end{center}
  %\caption{Grid near the film cooling hole: (a) of normal injection; (b) of compound angle $\beta$=300}
\label{fig18b}

\scalebox{1}{\includegraphics[trim= 2mm 0 2mm 0, clip=true,width=3.25in]{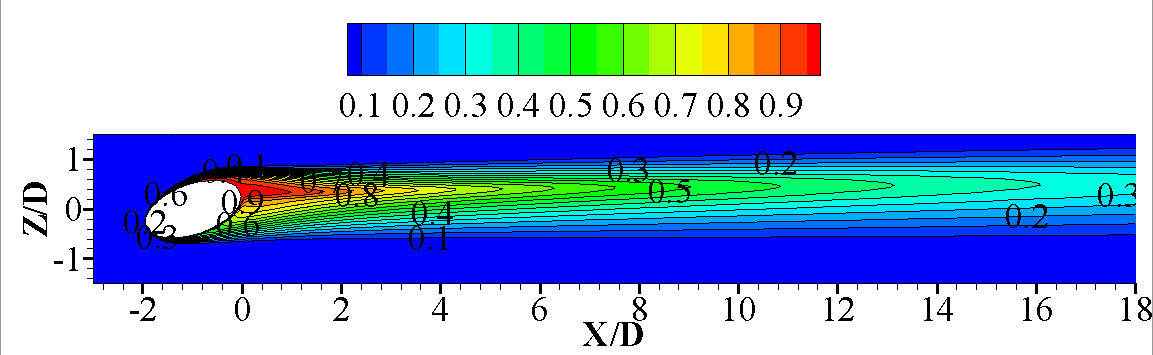}}
  %\centerline{\includegraphics{Fig6(a).eps}}% Images in 100% size
  \begin{center}
  (c)\\
  \end{center}
  %\caption{Grid near the film cooling hole: (a) of normal injection; (b) of compound angle $\beta$=300}
\label{fig18c}

\scalebox{1}{\includegraphics[trim= 2mm 0 2mm 0, clip=true,width=3.25in]{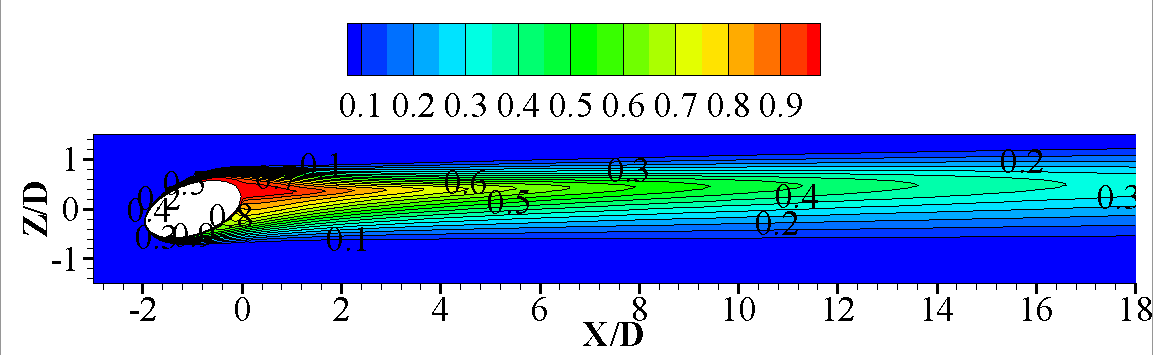}}
  %\centerline{\includegraphics{Fig6(a).eps}}% Images in 100% size
  \begin{center}
  (d)\\
  \end{center}
  \caption{Contours of adiabatic film cooling effectiveness over the plate for $\beta$ =$30\degree$ with: (a) L/D=1; (b) L/D=2; (c) L/D=3; (d) L/D=4 }
\label{fig18d}
\end{figure}

A major effect on the lateral spread is seen in this study due to compound holes that is as compared to normal injection hole, compound hole gives better lateral spread of cold stream over the plate. Fig. 18-20 shows that due to shape and the lateral vectoring of the cold stream the coverage area of the cold stream is very much improved. Rather seeing that the longitudinal spread is in control by L/D ratio as L/D=3 for compound angle $\beta$ =$60\degree$ provides better coverage area then L/D=4 with $\beta$ =$60\degree$ which can be seen in Fig. 20(c) and 20(d).

\begin{figure}
\scalebox{1}{\includegraphics[trim= 2mm 0 2mm 0, clip=true,width=3.25in]{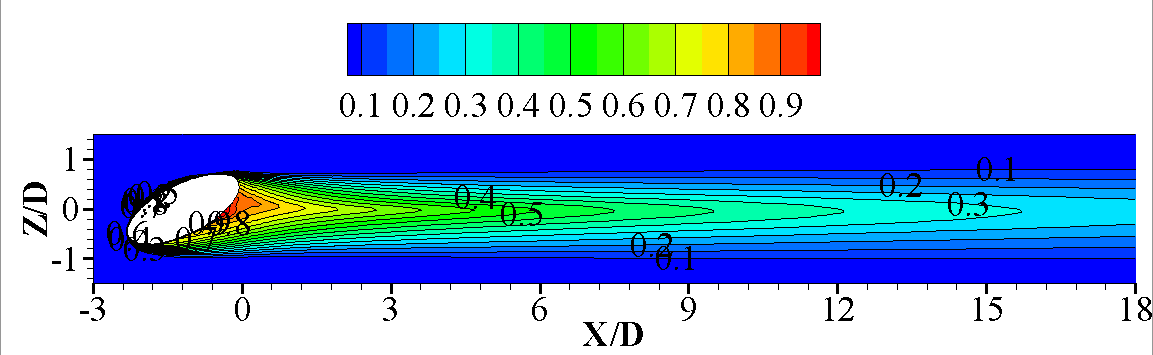}}
  %\centerline{\includegraphics{Fig6(a).eps}}% Images in 100% size
  \begin{center}
  (a)\\
  \end{center}
  %\caption{Grid near the film cooling hole: (a) of normal injection; (b) of compound angle $\beta$=300}
\label{fig19a}

\scalebox{1}{\includegraphics[trim= 2mm 0 2mm 0, clip=true,width=3.25in]{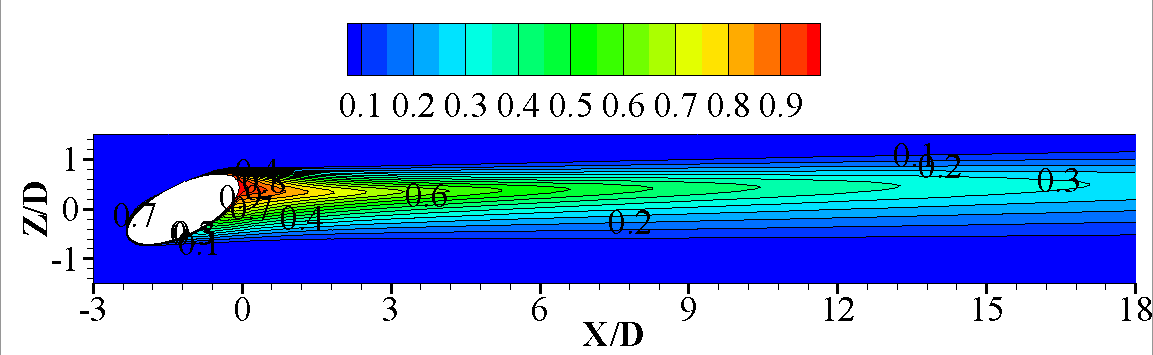}}
  %\centerline{\includegraphics{Fig6(a).eps}}% Images in 100% size
  \begin{center}
  (b)\\
  \end{center}
  %\caption{Grid near the film cooling hole: (a) of normal injection; (b) of compound angle $\beta$=300}
\label{fig19b}

\scalebox{1}{\includegraphics[trim= 2mm 0 2mm 0, clip=true,width=3.25in]{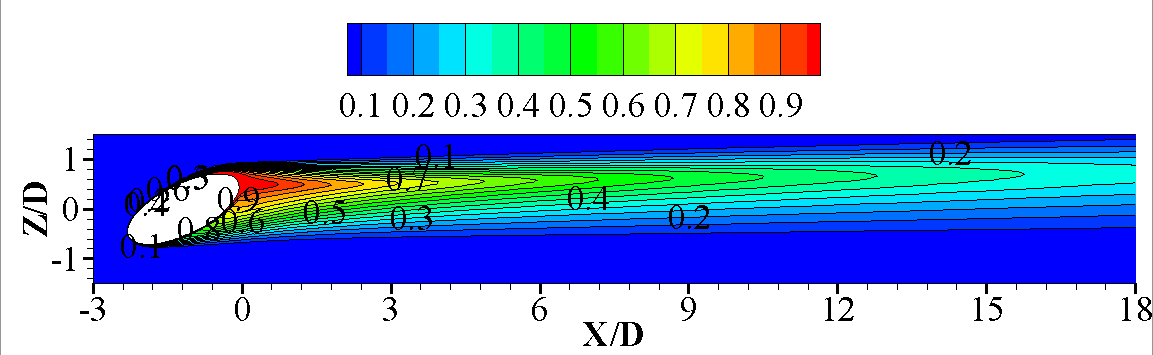}}
  %\centerline{\includegraphics{Fig6(a).eps}}% Images in 100% size
  \begin{center}
  (c)\\
  \end{center}
  %\caption{Grid near the film cooling hole: (a) of normal injection; (b) of compound angle $\beta$=300}
\label{fig19c}

\scalebox{1}{\includegraphics[trim= 2mm 0 2mm 0, clip=true,width=3.25in]{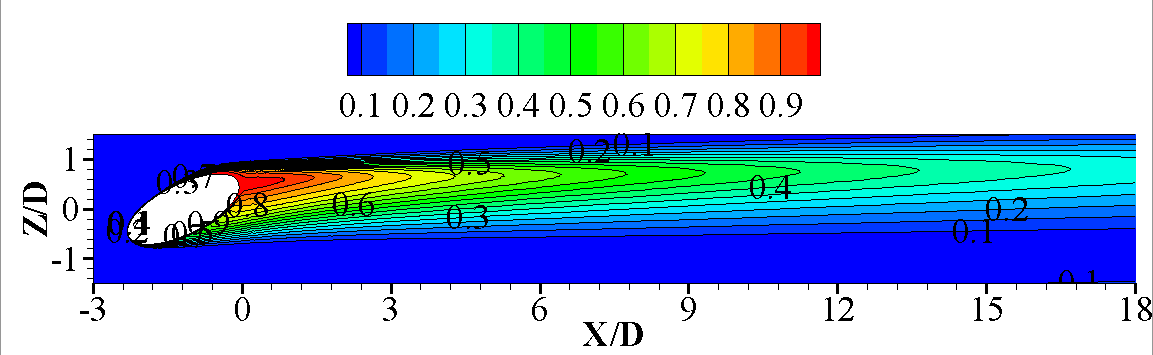}}
  %\centerline{\includegraphics{Fig6(a).eps}}% Images in 100% size
  \begin{center}
  (d)\\
  \end{center}
  \caption{Contours of adiabatic film cooling effectiveness over the plate for $\beta$ =$45\degree$ with: (a) L/D=1; (b) L/D=2; (c) L/D=3; (d) L/D=4 }
\label{fig19d}
\end{figure}

\begin{figure}
\scalebox{1}{\includegraphics[trim= 1mm 0 1mm 0, clip=true,width=3.25in]{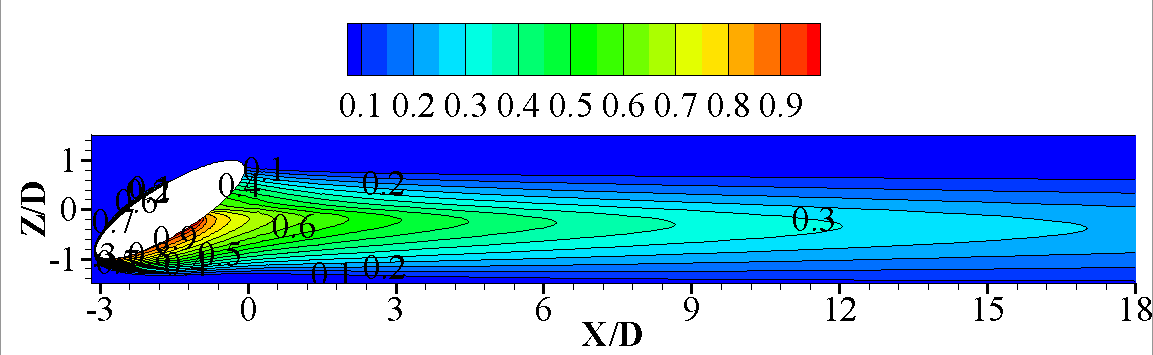}}
  %\centerline{\includegraphics{Fig6(a).eps}}% Images in 100% size
  \begin{center}
  (a)\\
  \end{center}
  %\caption{Grid near the film cooling hole: (a) of normal injection; (b) of compound angle $\beta$=300}
\label{fig20a}

\scalebox{1}{\includegraphics[trim= 1mm 0 1mm 0, clip=true,width=3.25in]{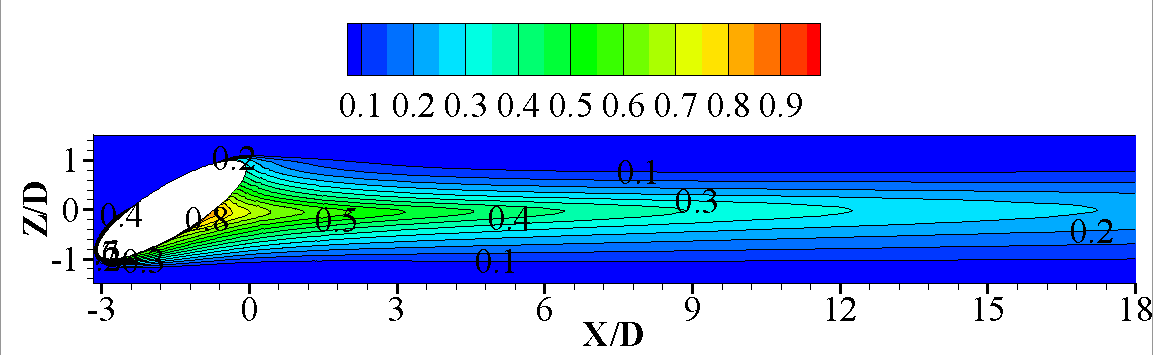}}
  %\centerline{\includegraphics{Fig6(a).eps}}% Images in 100% size
  \begin{center}
  (b)\\
  \end{center}
  %\caption{Grid near the film cooling hole: (a) of normal injection; (b) of compound angle $\beta$=300}
\label{fig20b}

\scalebox{1}{\includegraphics[trim= 1mm 0 1mm 0, clip=true,width=3.25in]{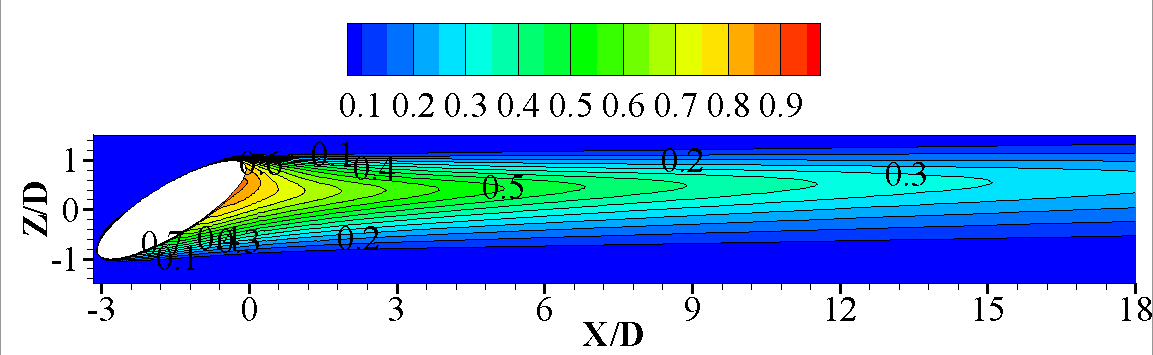}}
  %\centerline{\includegraphics{Fig6(a).eps}}% Images in 100% size
  \begin{center}
  (c)\\
  \end{center}
  %\caption{Grid near the film cooling hole: (a) of normal injection; (b) of compound angle $\beta$=300}
\label{fig20c}

\scalebox{1}{\includegraphics[trim= 1mm 0 1mm 0, clip=true,width=3.25in]{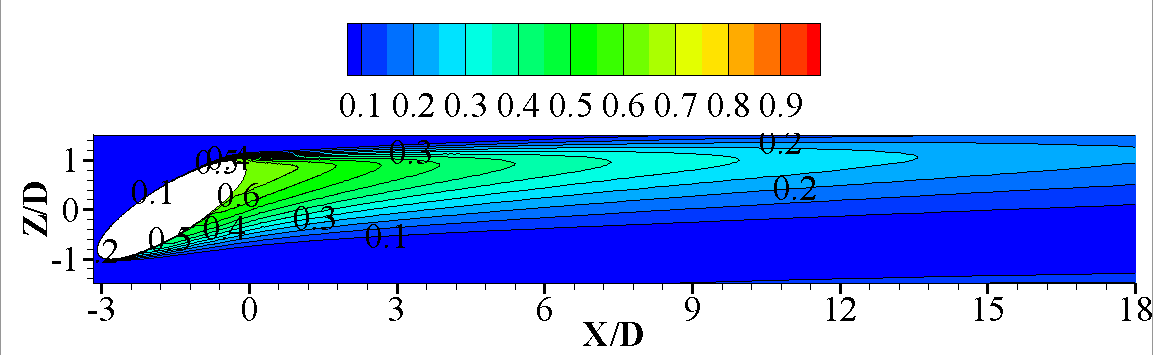}}
  %\centerline{\includegraphics{Fig6(a).eps}}% Images in 100% size
  \begin{center}
  (d)\\
  \end{center}
  \caption{Contours of adiabatic film cooling effectiveness over the plate for  $\beta$ =$60\degree$ with: (a) L/D=1; (b) L/D=2; (c) L/D=3; (d) L/D=4 }
\label{fig20d}
\end{figure}

{ \bf Adiabatic Film Cooling Effectiveness. }
It was assumed that the plate is adiabatic thus it is not participating in heat transfer due to conduction. The main purpose of present study is to study the effect of L/D ratio and compound angle on adiabatic film cooling effectiveness. \par
 
 \begin{figure}
\scalebox{1}{\includegraphics[width=3.25in]{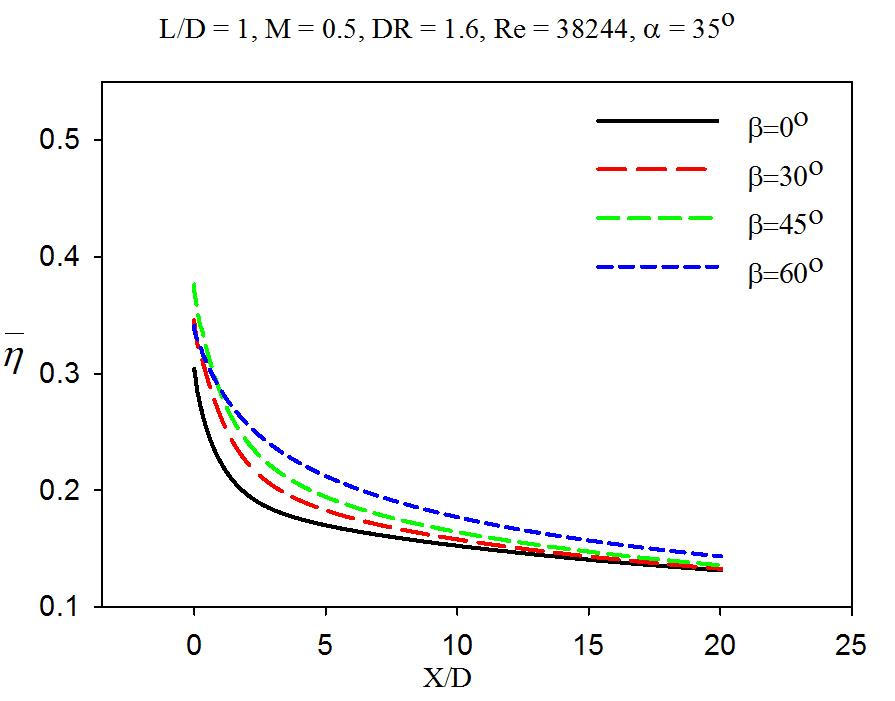}}
  %\centerline{\includegraphics{Fig6(a).eps}}% Images in 100% size
  \begin{center}
  (a)\\
  \end{center}
  %\caption{Grid near the film cooling hole: (a) of normal injection; (b) of compound angle $\beta$=300}
\label{fig21a}
\end{figure} 

\begin{figure}
\scalebox{1}{\includegraphics[width=3.25in]{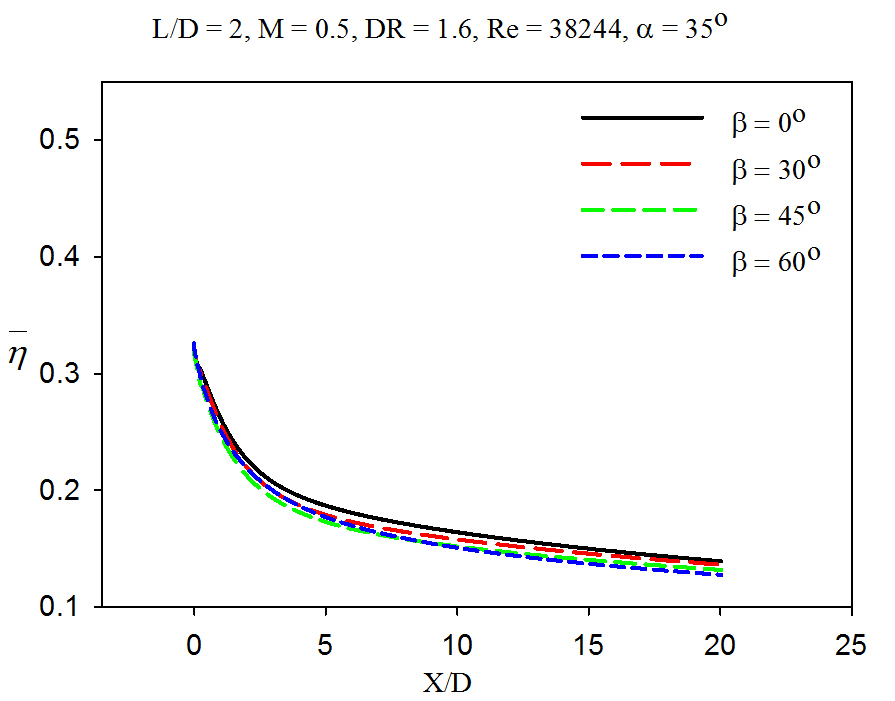}}
  %\centerline{\includegraphics{Fig6(a).eps}}% Images in 100% size
  \begin{center}
  (b)\\
  \end{center}
  %\caption{Grid near the film cooling hole: (a) of normal injection; (b) of compound angle $\beta$=300}
\label{fig21b}
\end{figure} 

\begin{figure}
\scalebox{1}{\includegraphics[width=3.25in]{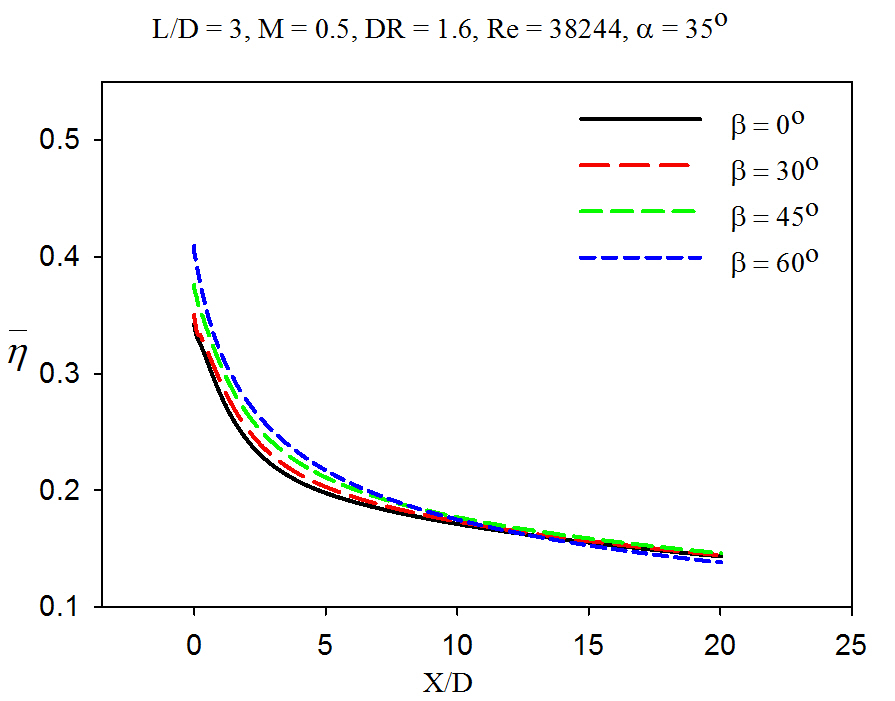}}
  %\centerline{\includegraphics{Fig6(a).eps}}% Images in 100% size
  \begin{center}
  (c)\\
  \end{center}
  %\caption{Grid near the film cooling hole: (a) of normal injection; (b) of compound angle $\beta$=300}
\label{fig21c}
\end{figure} 

\begin{figure}
\scalebox{1}{\includegraphics[width=3.25in]{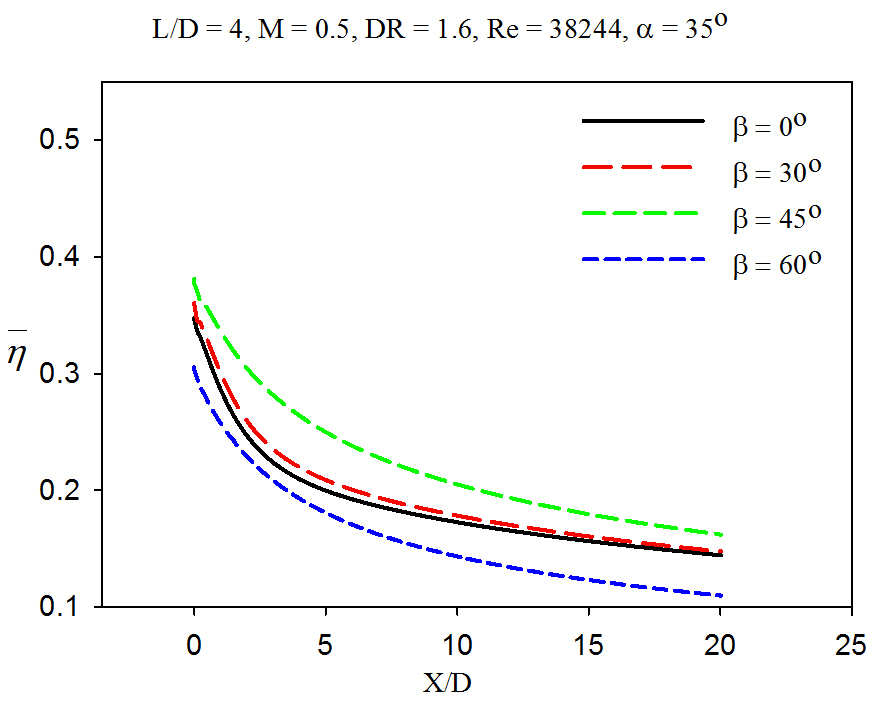}}
  %\centerline{\includegraphics{Fig6(a).eps}}% Images in 100% size
  \begin{center}
  (d)\\
  \end{center}
  \caption{Effects of compound angle on investigated L/D ratios}
\label{fig21d}
\end{figure}

Effect of the compound angles on laterally averaged adiabatic film cooling effectiveness is shown in Fig. 21. At L/D=1 higher compound angle gives better film cooling effectiveness and $\bar{\eta}60\degree > \bar{\eta}45\degree  >  \bar{\eta}30\degree > \bar{\eta}0\degree $.  At L/D = 2, difference in film cooling effectiveness at various compound angles is almost same. When L/D=3, average film cooling effectiveness follows same trend as that followed by L/D=1 case up to downstream distance X/D=10. After that average film cooling effectiveness of all the compound angles are almost same. At L/D=4, average film cooling effectiveness of $\beta = 60\degree$  is lowest while it increases as compound angle increases from $0\degree$ to $45\degree$ the effectiveness contours are also consistent with this observation  at compound angle $\beta = 60\degree$ which is due to the high mixing of cold and hot stream and can be seen in Fig. 20(d). \par
It can be seen from the Fig. 17-20 that with increase in compound angle increases surface coverage of secondary fluid. Same amount of secondary fluid is now dispersed over larger area. At L/D=1 dispersion of secondary fluid over plate increases as compound angle increases and hence film cooling effectiveness increases. At L/D=2 dispersion of secondary fluid is less as compared to L/D=1 but more compare to no compound angle in [3] and [14]. as shown in Fig. 17-20. It is also observed that with increase in compound angle at L/D=2 dispersion of coolant on plate is decreasing with very less deviation and a opposite trend has been seen as compared to L/D=1. At L/D=3 same trend has been seen like L/D=1, average film cooling effectiveness is increasing with increase of compound angle from $0\degree$ to $60\degree$ but the deviations in the magnitude of the effective is up to X/D=10 as shown in Fig. 21 \par

\begin{figure}
\scalebox{1}{\includegraphics[width=3.25in]{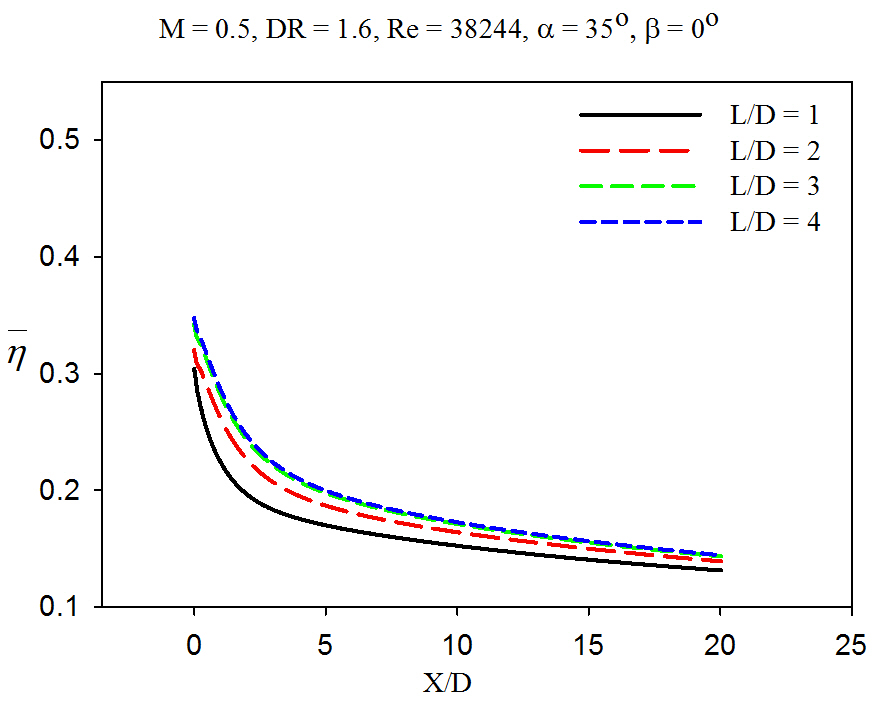}}
  %\centerline{\includegraphics{Fig6(a).eps}}% Images in 100% size
  \begin{center}
  (a)\\
  \end{center}
  %\caption{Grid near the film cooling hole: (a) of normal injection; (b) of compound angle $\beta$=300}
\label{fig22a}
\end{figure} 

\begin{figure}
\scalebox{1}{\includegraphics[width=3.25in]{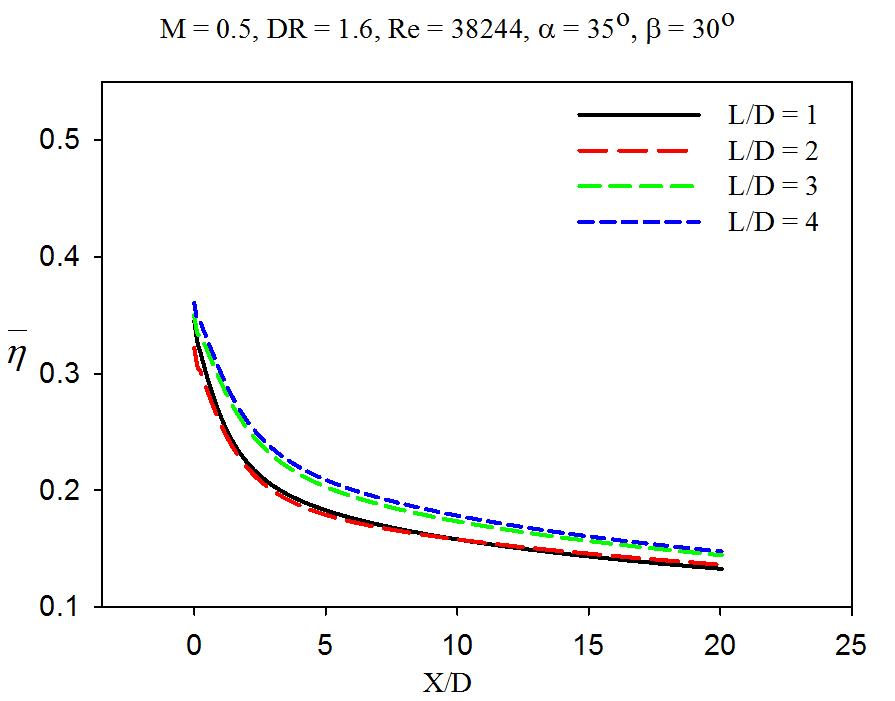}}
  %\centerline{\includegraphics{Fig6(a).eps}}% Images in 100% size
  \begin{center}
  (b)\\
  \end{center}
  %\caption{Grid near the film cooling hole: (a) of normal injection; (b) of compound angle $\beta$=300}
\label{fig22b}
\end{figure} 

\begin{figure}
\scalebox{1}{\includegraphics[width=3.25in]{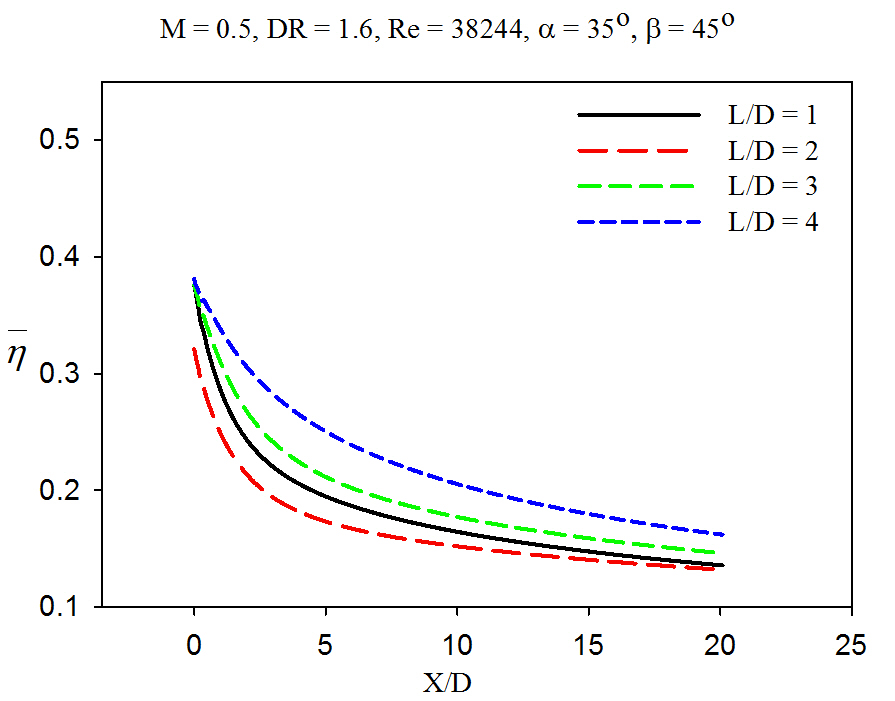}}
  %\centerline{\includegraphics{Fig6(a).eps}}% Images in 100% size
  \begin{center}
  (c)\\
  \end{center}
  %\caption{Grid near the film cooling hole: (a) of normal injection; (b) of compound angle $\beta$=300}
\label{fig22c}
\end{figure} 

\begin{figure}
\scalebox{1}{\includegraphics[width=3.25in]{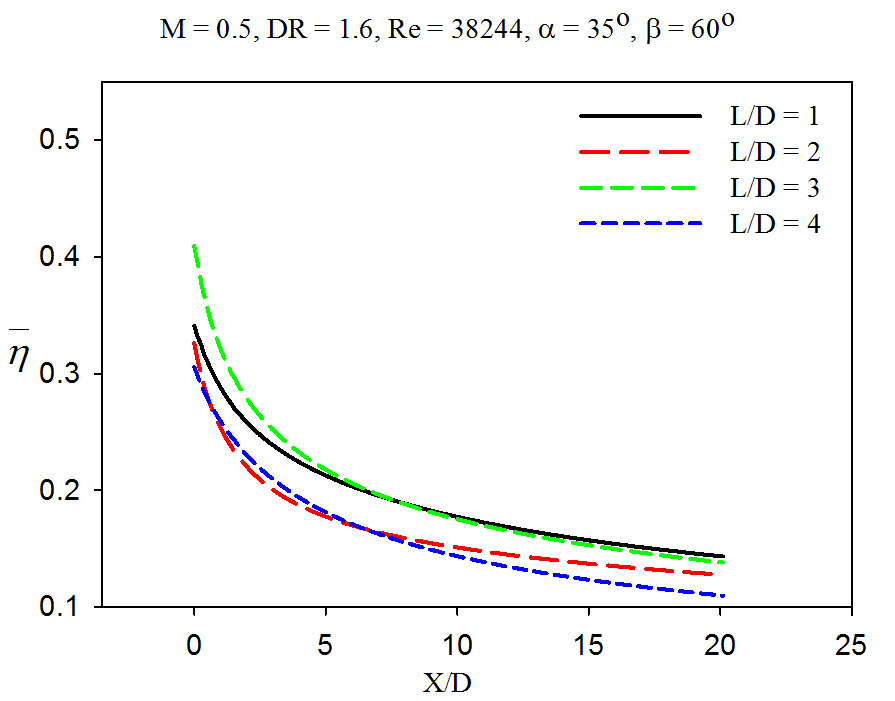}}
  %\centerline{\includegraphics{Fig6(a).eps}}% Images in 100% size
  \begin{center}
  (d)\\
  \end{center}
  
  \caption{Effect of L/D ratio for investigated compound angles }
\label{fig22d}
\end{figure}

%\begin{figure}
%  \centering
%  \mbox{
%    \subfigure[<(a)>\label{fig22d}]{\includegraphics[width=3.25in]{Fig22d.jpg}\quad
%    \subfigure[<(a)>\label{fig22d}]{\includegraphics[width=3.25in]{Fig22d.jpg}\quad
%    \subfigure[<(a)>\label{fig22d}]{\includegraphics[width=3.25in]{Fig22d.jpg}\quad
%    \subfigure[<(a)>\label{fig22d}]{\includegraphics[width=3.25in]{Fig22d.jpg}\quad
%  }
%  \caption{<Effect of L/D ratio for investigated compound angles>}
%  \label{fig22}
%\end{figure}

%\begin{center}
%
%\begin{table}
%\begin{tabular}{c c}
%%\textbf{blah}  & \textbf{text} $m=\pi,n=e$  \\
%\scalebox{1}{\includegraphics[width=3.25in]{Fig22a.jpg}} \text{A}&
%\scalebox{1}{\includegraphics[width=3.25in]{Fig22b.jpg}} \text{A}
% \\ 
%\scalebox{1}{\includegraphics[width=3.25in]{Fig22c.jpg}}\text{A}
% & 
%\scalebox{1}{\includegraphics[width=3.25in]{Fig22d.jpg}} \text{A}
%  \\
%%\textbf{text} & text
%% \caption{Pictures of animals}
%\end{tabular}
%\end{table}
%\end{center}
%

Effect of L/D ratio at investigated compound angles is shown in Fig. 22. At $0\degree$ compound angle, average film cooling effectiveness increase as L/D ratio increases from 1 to 4. At compound angle $\beta = 30\degree$, average film cooling effectiveness decreases as L/D increase from 1 to 2 and thereafter increases. Similar trend is followed at compound angle of $\beta = 45\degree$.  But at compound angle of $ 60\degree$, L/D=3 shows best effectiveness up to X/D=5, among all investigated L/D ratios. After this downstream distance, L/D=1 and 3 shows almost same and higher average film cooling effectiveness.  \par

Temperature contours shown in Fig. 17-20 also consists of these results. Temperature contours for $0\degree$ compound angle showing that the area covered by the secondary fluid over the plate has been stretched in downstream which leads to increase in average film cooling effectiveness with increase of L/D from 1 to 4. At compound angle $\beta = 30\degree$, from Fig. 17-20 it has been seen that the area covered by the secondary fluid starts moving towards left with increase in L/D. However, at L/D=2 it has been seen in the contours that larger area of plate is hot as compared to L/D=1 which leads to the sudden decrease in average film cooling effectiveness when L/D is increased to 2. At compound angle $45\degree$, similar trend has been seen as in compound angle $30\degree$ but the deviation in averaged film cooling effectiveness is large. From temperature contours Fig. 17-20 it is also clear that with increase in compound angle there is large shifting of secondary fluid stream in left direction with respect to mainstream velocity. \par
At compound angle $60\degree$, the trend shown in compound angle $30\degree$ and $45\degree$ has been followed for L/D=1, 2 and 3. And at L/D=4 there is sudden decrease in average film cooling effectiveness. It can be seen in Fig. 17-20 that there is similar shifting of secondary fluid stream on the plate as in the case of $30\degree$ and $45\degree$ compound angle. It was worth noting that it is clear that the film cooling effectiveness of higher compound angle with higher L/D ratio, such as happening in this study with $\beta = 60\degree$ and L/D = 4 is lower with significant number. 
\par

%%%%%%%%%%%%%%%%%%%%%%%%%%%%%%%%%%%%%%%%%%%%%%%%%%%%%%%%%%%%%%%%%%%%%%
\section*{Conclusions}
Investigation done in the present study shows various effects of L/D ratio and compound angle on film cooling effectiveness.\par
Increase in compound angle results in increase in lateral coverage area of secondary cold fluid and dissipation of one side vortex due to lateral velocity of cold flow. L/D ratio effects the flow pattern in an effective way to increase coverage of cold stream in downstream. Film cooling effectiveness of compound angled hole is better than simple hole for all investigated L/D ratio except L/D=2, where simple injection hole has marginal higher effectiveness. Higher L/D ratio with higher compound angle is not a good configuration for better cooling effectiveness as it  increases mixing of cold stream and hot stream. \par
Compound angle $45\degree$, as concluded from the results, performed very well for all L/D ratios and as compared to higher compound angle it has shown very significant increases in averaged film cooling effectiveness.\par
Performance of L/D=4 is very well appreciated for all compound holes except $\beta$ =$60\degree$, and specifically for compound angle $\beta$ =$45\degree$ the film cooling effectiveness is better. While compound angle $\beta$ =$60\degree$ performed better than any other compound angle at L/D=3 for small X/D in downstream.

%%%%%%%%%%%%%%%%%%%%%%%%%%%%%%%%%%%%%%%%%%%%%%%%%%%%%%%%%%%%%%%%%%%%%%

\end{document}